\documentclass[aps,prb,10pt,twocolumn,floatfix,showpacs,superscriptaddress,longbibliography]{revtex4-1}
\usepackage[colorlinks=true,citecolor=blue,linkcolor=blue,urlcolor=blue]{hyperref}
\usepackage[normalem]{ulem}
\usepackage{amsmath}
\usepackage{amssymb}
\usepackage{graphicx,amsmath,amsfonts,bm}
\usepackage{epstopdf}
\usepackage{bm}
\usepackage{bbm}
\usepackage{color}
\usepackage{relsize}
\usepackage{dsfont}
\usepackage{braket}
\usepackage[caption=false]{subfig}
\usepackage{hyperref}
\usepackage[all]{hypcap} 
\usepackage[T1]{fontenc}
\usepackage{dsfont}
\usepackage{mathbbol}

\allowdisplaybreaks

\begin{document}
\title{Limits of Thermal Conductance Quantization in Chiral Topological Josephson Junctions}

\author{Daniel~Gresta}
\thanks{These authors contributed equally to this work.}
\affiliation{Institute for Theoretical Physics and Astrophysics and W{\"u}rzburg-Dresden Cluster of Excellence ct.qmat,\\
Julius-Maximilians-Universit{\"a}t W{\"u}rzburg, D-97074 W{\"u}rzburg, Germany}

\author{Fernando~Dominguez}
\thanks{These authors contributed equally to this work.}
\affiliation{Institute for Theoretical Physics and Astrophysics and W{\"u}rzburg-Dresden Cluster of Excellence ct.qmat,\\
Julius-Maximilians-Universit{\"a}t W{\"u}rzburg, D-97074 W{\"u}rzburg, Germany}

\author{Florian~Goth}
\affiliation{Institute for Theoretical Physics and Astrophysics and W{\"u}rzburg-Dresden Cluster of Excellence ct.qmat,\\
Julius-Maximilians-Universit{\"a}t W{\"u}rzburg, D-97074 W{\"u}rzburg, Germany}

\author{Raffael~L.~Klees}
\affiliation{Institute of Physics, University of Augsburg, D-86159 Augsburg, Germany}

\author{Laurens W. Molenkamp}
\affiliation{Experimental Physics III, Julius-Maximilians-Universit\"at W\"urzburg, D-97074 W\"urzburg, Germany}
\affiliation{Institute for Topological Insulators, Am Hubland, 97074 W\"urzburg, Germany}

\author{Ewelina~M.~Hankiewicz}
\affiliation{Institute for Theoretical Physics and Astrophysics and W{\"u}rzburg-Dresden Cluster of Excellence ct.qmat,\\
Julius-Maximilians-Universit{\"a}t W{\"u}rzburg, D-97074 W{\"u}rzburg, Germany}

\begin{abstract}
We investigate thermal and  non-local electrical  transport in four-terminal Josephson junctions formed by a normal region coupled to two transverse chiral superconducting leads, supporting phases characterized by Chern numbers ${\cal C}=0,\,1$\,and\,2. 
We identify the conditions under which a single chiral Majorana mode (${\cal C}=1$) produces a robust half-quantized thermal conductance, while non-local electrical conductance remains strongly suppressed by particle-hole symmetry. Thermal conductance quantization occurs near a superconducting phase difference $\pi$, but only in the low-doping regime of the central region and in the intermediate- to long-junction limits.
At finite Zeeman fields, the thermal response broadly follows the topology of the isolated superconducting leads for the $C=1$ phase while, in the ${\cal C}=2$ phase, 
the thermal conductance generally deviates from quantization, depending on the momentum-space location of the Majorana modes.
Our results establish clear criteria for probing chiral Majorana modes in Josephson junctions and highlight the essential role of momentum-space structure, finite-size geometry, and sample parameters in thermal transport.
\end{abstract}

\date{\today}
\maketitle
\section{Introduction}\label{Sec:Intro}

Majorana bound states (MBSs) are exotic quasiparticle excitations in condensed matter systems whose creation and annihilation operators are identical, $\gamma=\gamma^\dagger$ \cite{Majorana1937a, Read2000a, Ivanov2001a, Kitaev2001a}. As a consequence, they carry no net charge and do not possess a well-defined occupation number, endowing them with properties that are fundamentally distinct from those of conventional fermionic excitations. 

\begin{figure}[ht!]
    \centering
    \includegraphics[width=0.75\linewidth]{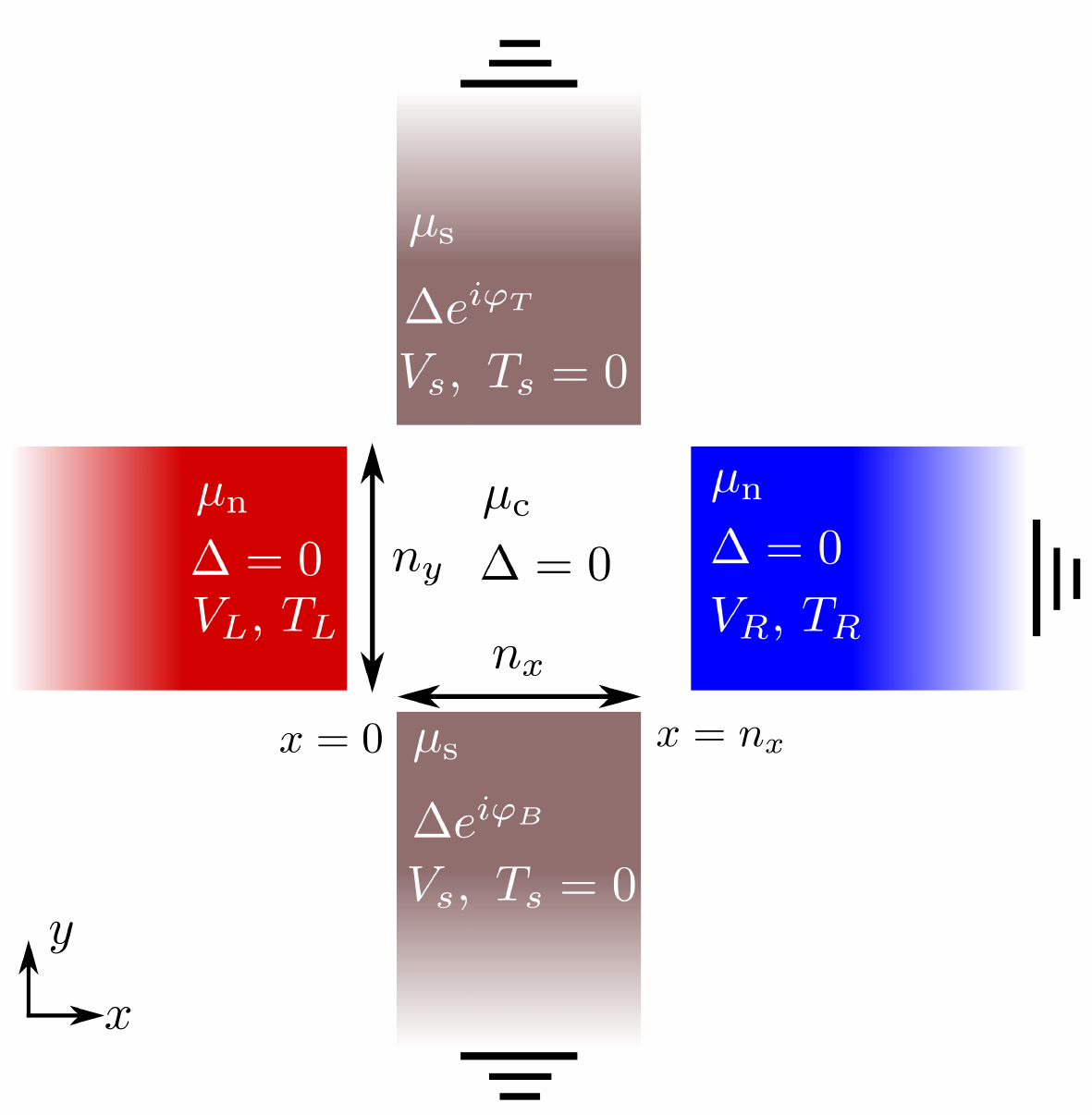}
    \caption{Schematic of the four-terminal Josephson junction. A normal region of size $n_x \times n_y$ and chemical potential $\mu_c$ is coupled to two normal leads (left $L$ and right $R$) and two superconducting leads (top $T$ and bottom $B$). The superconducting terminals are grounded, share a common chemical potential $\mu_s$, and are kept at temperature $T_s$, with pairing potentials $\Delta e^{i\varphi_{T,B}}$. The right normal lead is also grounded ($V_R=0$), while the left normal lead is biased by a voltage $V_L$ and a temperature difference $T_L$, providing the sole source of nonequilibrium. Black bars indicate grounded terminals.
    } 
    \label{fig:set-up}
\end{figure}

Among the various candidate systems, $p$-wave superconductors are the most extensively studied platforms for realizing topological superconductivity~\cite{Volovik1999a, Kitaev2001a}. However, the apparent absence of this superconducting phase in nature has compelled researchers to engineer such a superconducting phase by the combination of spin-orbit coupling, magnetic fields and superconductivity~\cite{Perge2014a,Jeon2017a, Fu2008a, Fu2009a, Lutchyn2010a, Oreg2010a, Penaranda2023a, San-Jose2015a,Finocchiaro2018a, Suominen2017a, Nichele2017a,Ribeiro_2022,Ribeiro_2023, traverso2025a,Alicea2012a}.

A variety of experimental signatures have been proposed to detect MBSs, most notably, the quantized zero-bias conductance peak $G=2e^2/h$ in normal–superconductor junctions \cite{Law2009a} and the fractional Josephson effect characterized by a $4\pi$-periodic current–phase relation \cite{Kitaev2001a}. However, experimental observations often deviate from ideal theoretical predictions: Zero-bias conductance peaks are frequently found to be suppressed or non-quantized~\cite{Mourik2012a,Deng2012a, Das2012a,Nichele2017a}, while the fractional Josephson effect occurs only in a restricted region of the parameter space of the devices tested
~\cite{Rokhinson2012a, Wiedenmann2016a,Li2018a,Bocquillon2016a, Deacon2017a,Laroche2019a}.
Surprisingly, experimental signatures of the fractional Josephson effect have been observed at zero magnetic field, i.e.~in the presence of time-reversal symmetry, where emission to the quasicontinuum should take place
~\cite{Wiedenmann2016a,Li2018a,Bocquillon2016a,Deacon2017a,Ruiz2024a,Laroche2019a}. Although some works suggest possible mechanisms to recover an effective 4$\pi$-periodicity without introducing a magnetic field~\cite{Sticlet2018a,heinz2024a}, the most likely scenario that gives rise to the experimental observations is the presence of a parasitic impedance~\cite{Liu2025}.

Additionally, the presence of trivial zero-energy excitations makes the unambiguous detection of these exotic quasiparticles more complicated~\cite{Prada2012a,Liu2017a,Fleckenstein2018a, Moore2018a, Marra2019a,Dmytruk2020a,Mateos2024a,Oladunjoye2019a}. Hence, recent efforts go in the direction of designing detection schemes that enable distinguishing between trivial and topological scenarios~\cite{Haim2015a, Fleckenstein2021a, Pakizer2021a, Dominguez2024a, Bittermann2024a}. In this context, thermoelectric measurements have emerged as a powerful complementary tool~\cite{Akhmerov2011a, Wang2011a, Bauer2021a, Ning-Xuan2022a, Arrachea2025a,Sturm2025a,Lopez2014a}. Indeed, Majorana excitations emerging in topological Josephson junctions (JJ) at a superconducting phase difference $\phi = \pi$—in close analogy with the zero-energy Jackiw–Rebbi modes—can be directly probed through their thermal transport signatures: A single Majorana bound state carries only half of the degrees of freedom of a conventional fermionic mode, a fact encoded in its central charge $c=1/2$~\cite{Wang2011a}. 
Consequently, when such a mode dominates heat transport, the thermal conductance reaches the universal quantized value, $\kappa = 0.5\,\kappa_0$, with $\kappa_0 = \pi^2 k_B^2 T/(3h)$.

In the present paper, we investigate the geometrical and energy conditions for thermal conductance quantization in a topological Josephson junction. To this aim, we construct a Josephson junction setup with superconducting leads that can host a topological phase with either zero, one, or two chiral Majorana modes, and probe the corresponding Andreev bound states by attaching normal leads to the central part of the junction; see Fig.~\ref{fig:set-up}. In this setup, we study the interplay of the Thouless energy $E_T$, the superconducting gap $\Delta$, and the doping of the central part of the junction. 

The paper is organized as follows. In Sec.~\ref{sec.:Model}, we introduce the theoretical framework of the work, including the continuum Bogoliubov-de Gennes(BdG) Hamiltonian and the four-terminal geometry used to probe non-local transport. In Sec.~\ref{sec.Topolead}, we analyze the topological phase diagram of the superconducting Hamiltonian. 
In Sec.~\ref{Sec.JJZero}, we focus on the zero-Zeeman-field regime, which supports a chiral Majorana mode, with ${\cal C}=1$. Here, we study the Andreev bound states, and the conditions needed to observe the half-quantized thermal conductance $\kappa = 0.5\kappa_0$, while the non-local electrical conductance remains suppressed. In Sec.~\ref{Sec.JJFinite}, we extend the analysis to finite Zeeman field, where the superconducting leads can host a ${\cal C}=2$ topological phase. 
Finally, Sec.~\ref{Sec.Conclusions} summarizes our results and discusses their implications.

\section{Model and formalism.}\label{sec.:Model}
In this section, we introduce the bulk Hamiltonian used to build the four terminal junction and present the transport formalism based on non-equilibrium Green's functions. 

\subsection{Model Hamiltonian}\label{sec. continum Ham}
We model the superconducting leads using the lattice-regularized Dirac-Bogoliubov-de Gennes Hamiltonian introduced in Ref. \onlinecite{qi2010chiral},
\begin{align}
    \label{eq:hamiltonian}
    {\cal H} =  & \sum_{\alpha=x,y} t \sin(k_\alpha a) \sigma_\alpha \otimes \tau_z   + [\mathcal{Z}+m(\mathbf{k})] \sigma_z\otimes \tau_0
    \nonumber \\
    & -\mu\sigma_0 \otimes \tau_z + \Delta \sigma_0 \otimes \tau_\varphi,
\end{align}
where $\boldsymbol{\sigma}$($\boldsymbol{\tau}$) are Pauli matrices acting in spin (particle-hole) space, and $t= \hbar v_\text{F}/a$ sets the spin-orbit energy scale with $v_\text{F}$ the Fermi velocity and $a$ the lattice constant. 
The parameters $\mu$, $\Delta$, and $\mathcal{Z}$ denote the chemical potential, the superconducting pairing amplitude and the applied Zeeman field, respectively. The momentum-dependent mass term is given by $m(\mathbf{k})=2m_0[2-\cos(k_x a)-\cos(k_y a)]$, with $m_0$ the Wilson mass.
We introduce the matrix $\tau_\varphi=\cos(\varphi)\tau_x+\sin(\varphi)\tau_y$ to incorporate a superconducting phase $\varphi$, corresponding to a pairing potential $\Delta e^{i\varphi}$; this formulation will be useful when constructing the Josephson junction. Throughout this work, energies are measured in units of $t$.

\begin{figure}[tb]
    \centering
    \includegraphics[width=1.0\linewidth]{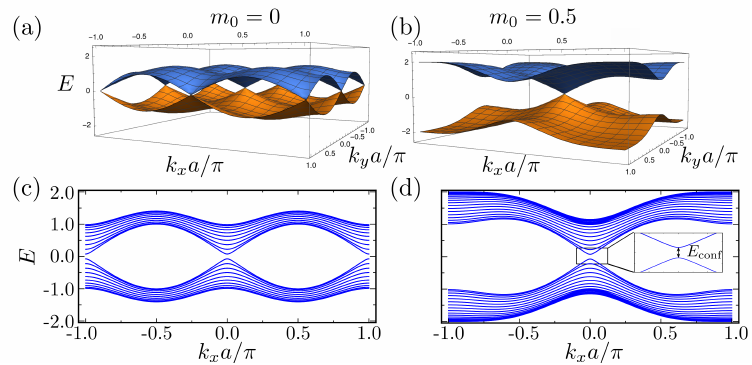}
    \caption{Bulk energy spectrum of the normal ($\Delta=0$) Hamiltonian, with $\mathcal{Z}=0$, $\mu=0$ and two different values of the Wilson mass, $m_0=0$ (a) and $m_0=0.5$ (b). Energy spectrum for a nanoribbon with $n_y=20a$, also for $m_0=0$ (c) and $m_0=0.5$ (d). All panels are in units of $t$.}
    \label{fig:bulkbands}
\end{figure}

For zero Zeeman field $\mathcal{Z}=0$ and in the long-wavelength limit $k_\alpha a\rightarrow 0$, Eq.~\eqref{eq:hamiltonian} reduces to 
\begin{align}\label{eq:approx}
\mathcal{H}\approx \hbar v_\text{F} \boldsymbol{k}\cdot \boldsymbol{\sigma}-\mu\sigma_0 \otimes \tau_z + \Delta \sigma_0 \otimes \tau_\varphi,
\end{align}
which corresponds to the Bogoliubov de Gennes Hamiltonian
of a proximitized topological insulator surface state \cite{Fu2008a}. Despite this formal similarity, important differences arise at finite momentum. To have more insight, we first analyze the normal-state spectrum by setting $\Delta=0$ and $\mathcal{Z}=0$ as shown in Fig.~\ref{fig:bulkbands} (a) and (b). 
In the absence of a Wilson mass ($m_0=0$), the energy spectrum exhibits four Dirac cones within the first Brillouin zone, placed at the high symmetry points, i.e.~$(k_x,k_y)=(0,0)$, $(0,\pi)$, $(\pi,0)$ and $(\pi,\pi)$.
Then, a finite Wilson mass $(m_0\neq 0)$ gaps out all Dirac cones located at finite momentum, resulting in a single Dirac cone centered at $(k_x,k_y)=(0,0)$ and its 
low-energy description is given by Eq.~\eqref{eq:approx}. The Wilson mass term explicitly breaks time-reversal symmetry, consequently, the superconducting phases from this Hamiltonian host chiral, rather than, helical Majorana modes, expected for a proximitized topological insulator surface state~\cite{Fu2008a}.
Nevertheless, the conditions to observe a quantized thermal conductance studied in this work are also relevant to helical platforms because as we will see, they depend on the number of Dirac cones.

Imposing hard-wall boundary conditions leads to a confinement-induced gap, $E_\text{conf}$, at the Dirac points. To see this, we consider a ribbon geometry that is finite in the $y-$ direction, with width $n_y$, and translationally invariant along $x$, so that $k_x$ remains a good quantum number, see Fig.~\ref{fig:bulkbands}(c,d). In this geometry, the low-energy bands at $k_x=0$ are quantized by transverse confinement and scale with the Thouless energy as $E_\text{conf}\propto (n+1/2)E_T$ where $n\in\mathbb{Z}$ and $E_T = \hbar v_\text{F}/n_y$. This energy scale plays a central role throughout this work, as nonlocal transport requires propagating modes connecting the left and right normal leads. Such modes are present when $E_\text{conf}\ll\Delta$, corresponding to the intermediate-to long-junction regime ~\cite{Titov2007a}.

\subsection{Transport Formalism}\label{sec. SNSintro}

We build a four-terminal junction according to Fig.~\ref{fig:set-up}, by bringing the Hamiltonian into its discrete version and set values for $\mu$ and $\Delta$, depending on the region: 
The normal regions of the junction and the transverse normal leads are obtained by setting $\Delta=0$ and ${\mathcal{Z}}=0$ in Eq.~\eqref{eq:hamiltonian} and $\mu=\mu_n$, for the normal leads and $\mu=\mu_\text{c}$ for the normal central region, with dimensions $n_x$ and $n_y$ in $x$- and $y$-direction, respectively. 
For the superconducting leads we use $\Delta=0.35$, $\mu_s$ and $\mathcal{Z}$, with the superconducting phase difference $\phi = \varphi_T-\varphi_B$ between the top $(T)$ and $(B)$ bottom leads.

In this setup, transport is probed between the normal leads: a voltage (or temperature) bias is applied to $L$, and the resulting current is measured at $R$, keeping all other leads, normal and superconducting, grounded.
In the linear-response regime, the non-local electrical conductance between terminals $L$ and $R$ is given by 
\begin{equation}
    \label{eq:def G}
    G = G_0 \int_{-\infty}^\infty d\varepsilon \left(-\frac{\partial f}{\partial \varepsilon} \right) {\cal T^E}, \qquad G_0 = \frac{e^2}{h}.
\end{equation}

Similarly, the thermal conductance at temperature $T$ is (neglecting thermoelectrical effects)
\begin{equation}
    \label{eq:def K}
    \kappa = \kappa_0 \int_{-\infty}^\infty d\varepsilon \left(\frac{\varepsilon}{k_B T}\right)^2\left(-\frac{\partial f}{\partial \varepsilon}\right){\cal T^Q},  ~\kappa_0 = \frac{\pi^2 k_B^2 T}{3h}.
\end{equation}

In the zero temperature limit $T\rightarrow0$, Eqs.~\eqref{eq:def G} and~\eqref{eq:def K} reduce to $G/G_0=\mathcal{T}^\mathcal{E}$ and  $\kappa/\kappa_0=\mathcal{T}^\mathcal{Q}$, with 
\begin{align}
    \label{TE}
    &{\cal T^E} = {\cal T}_{\mathrm{RL}}^{ee}-{\cal T}_{\mathrm{RL}}^{he}, \\
    &{\cal T^Q} = {\cal T}_{\mathrm{RL}}^{ee}+{\cal T}_{\mathrm{RL}}^{he},
\end{align}
and the transmission function 
\begin{equation}
    \label{eq:Caroli}
    {\cal T}_{RL}^{\alpha\beta}(\varepsilon)=\mathrm{Tr} \left[\Gamma_R {\cal G}^{r;\alpha \beta}_{RL} \Gamma_L {\cal G}^{a;\beta\alpha}_{LR}\right],   
\end{equation}
associated with a quasiparticle scattering from type $\beta \in \left\{ e,h\right\}$ from  $L=(0,y)$ to a type $\alpha$ in $R=(n_x,y)$~\cite{caroli1971direct}. Here, $\Gamma_k$ is the coupling rate to $k$-lead, and the trace is taken over all other degrees of freedom, i.e.:~spin and lattice position in $y$-direction. Moreover, $\mathcal{G}_{RL}^{r/a;\alpha\beta}(\varepsilon)$ is the retarded/advanced Green's function of the central part, evaluated at positions $R$ and $L$, see further details in App.~\ref{App.GF}. 

The local spectral properties of the junction are obtained directly from the Green’s function. The local density of states (LDOS) at energy $E$ and lattice position position $\boldsymbol{x}$ is given by
\begin{equation}
    \text{LDOS}(E,\boldsymbol{x})=\frac{1}{\pi} \text{Im}\text{Tr}\{{\cal G}_{\boldsymbol{x},\boldsymbol{x}}^{a}(E)\},
\label{eq:LDOS}
\end{equation}
where the trace runs over spin and particle–hole degrees of freedom. The total density of states (DOS) of the central region follows by summing the LDOS over all lattice sites within the normal part,

\begin{equation}
\label{eq:DOS}
\text{DOS}(E)
=\sum_{x=0}^{n_x}\sum_{y=0}^{n_y}
\text{LDOS}(E,\boldsymbol{x}).
\end{equation}

\section{Topological phase diagram of the Hamiltonian}\label{sec.Topolead}
In this section, we analyze the topological phases of the superconducting Hamiltonian given in Eq.~\eqref{eq:hamiltonian}, as a function of the model parameters, i.e.~$\mu_s$, ${\cal Z}$, $\Delta$ and $m_0$. 

The presence of either a finite Wilson mass ($m_0$) or a Zeeman field ($\mathcal{Z}$), breaks time-reversal symmetry setting the Hamiltonian in symmetry class D \cite{chiu2016classification}.
Then, according to the tenfold classification, in two dimensions, this Hamiltonian allows for a topological invariant $\mathbb{Z}$, the Chern number
\begin{equation}
    {\cal C} = \frac{1}{2\pi}\int_{BZ} F_{xy}(\mathbf{ k})  d\mathbf{ k}, 
\end{equation}
with $F_{xy}({\bf k})=\partial_x A_y-\partial_y A_x$, the Berry curvature and $A_\alpha= i\langle u_{\bf k}|\partial_\alpha| u_{\bf k}\rangle$, the Berry connection of the Bloch band $u_{\bf k}$ and the integration performed over the Brillouin-zone. 

\begin{figure}[tb]
    \centering
    \includegraphics[width=1.0\linewidth]{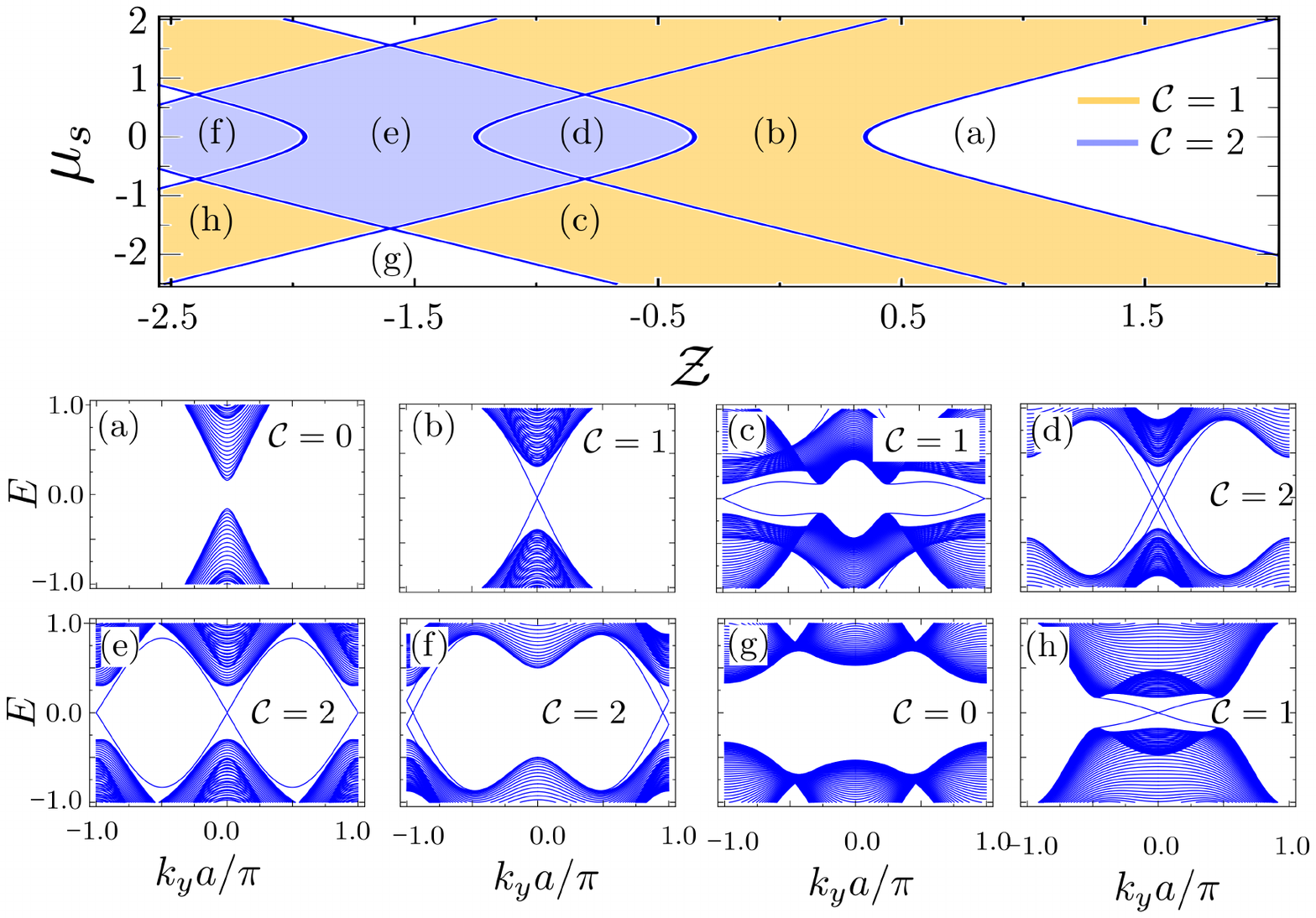}
        \caption{Top panel: Phase diagram of the bulk Hamiltonian as a function of $\mu_s$ and $\mathcal{Z}$, with $m_0=0.8$ and $\Delta=0.35$, delimited by blue curves marking gap closings at high symmetry points. Different colors, white, yellow and blue set $\mathcal{C}=0,\,1,\,2$, respectively. Panels (a)-(h) show exemplary energy spectra of a nanoribbon, with a width of $n_x=60a$. (a) $(\mathcal{Z}, \mu_s)$, (a) $(0.5,0.0)$, (b) $(0.0,0.0)$, (c) $(-0.5,1.5)$, (d) $(-0.75,0.2)$, (e) $(-1.7,0.2)$, (f) $(-2.5,0.2)$, (g) $(-1.5,2.0)$ and (h) $(-2.0,2.0)$.}
    \label{fig:phase diagram}
\end{figure}

\subsection{Hamiltonian at zero Zeeman Field}
We begin by considering the case of vanishing Zeeman field, $\mathcal{Z}=0$, which provides a useful reference point for understanding the structure of the phase diagram in Fig.~\ref{fig:phase diagram}. In this limit, time-reversal symmetry is broken solely by the Wilson mass $m_0$, allowing the system to access topologically-nontrivial phases even in the absence of an external magnetic field. This regime is particularly transparent, as it isolates the role of band inversion and momentum-space gap closings without the additional complexity introduced by Zeeman splitting.

The energy spectrum of the superconducting Hamiltonian in Eq.~\eqref{eq:hamiltonian} exhibits gap closings at high-symmetry points of the Brillouin zone, namely $(k_x,k_y)=(0,0)$, $(0,\pi)$, $(\pi,0)$, and $(\pi,\pi)$. These gap closings and subsequent reopenings delimit distinct topological phases characterized by different values of the Chern number.

For $|m_0|\leq \sqrt{\Delta^2+\mu_s^2}/4$, the system remains fully gapped and topologically-trivial, with $\mathcal{C}=0$. Upon increasing $|m_0|$, a topological phase transition occurs and the system enters a chiral superconducting phase with $\mathcal{C}=1$, supporting a single chiral Majorana edge mode.

Within the $\mathcal{C}=1$ phase, two distinct regimes can be identified,
\begin{align}
\sqrt{\Delta^2+\mu_s^2}/4 \leq &|m_0|\leq \sqrt{\Delta^2+\mu_s^2}/2,  \label{eq:topsector1}\\
|m_0|>&\sqrt{\Delta^2+\mu_s^2}/2, \label{eq:topsector2}
\end{align}
which differ by the momentum at which the gap closes and the Majorana mode crosses zero energy. In the former case, the zero-energy crossing occurs at finite momentum, $k_{x/y}=\pi/a$, whereas in the latter it takes place at the Brillouin-zone center, $k_{x/y}=0$.

\subsection{Finite Zeeman Field}

Without loss of generality\footnote{Although we do not explore the full parameter space, the chosen parameters allow us to access all topological phases supported by the Hamiltonian.}, we fix the Wilson mass to $m_0=0.8$ and the superconducting pairing amplitude to $\Delta=0.35$, and analyze the topological phases as functions of the Zeeman field $\mathcal{Z}$ and the chemical potential $\mu_s$. In this regime, bulk gap closings and reopenings delimit three distinct topological phases characterized by $\mathcal{C}=0,\,1,$ and $2$, as indicated with colored regions in the top panel of Fig.~\ref{fig:phase diagram}.
Alongside the bulk phase diagram, we show representative energy spectra of the BdG Hamiltonian for a nanoribbon geometry, with translational invariance along the $y$-direction (good quantum number $k_y$) and finite extent along $x$-direction.

\begin{figure}[tb]
    \centering
    \includegraphics[width=1.0\linewidth]{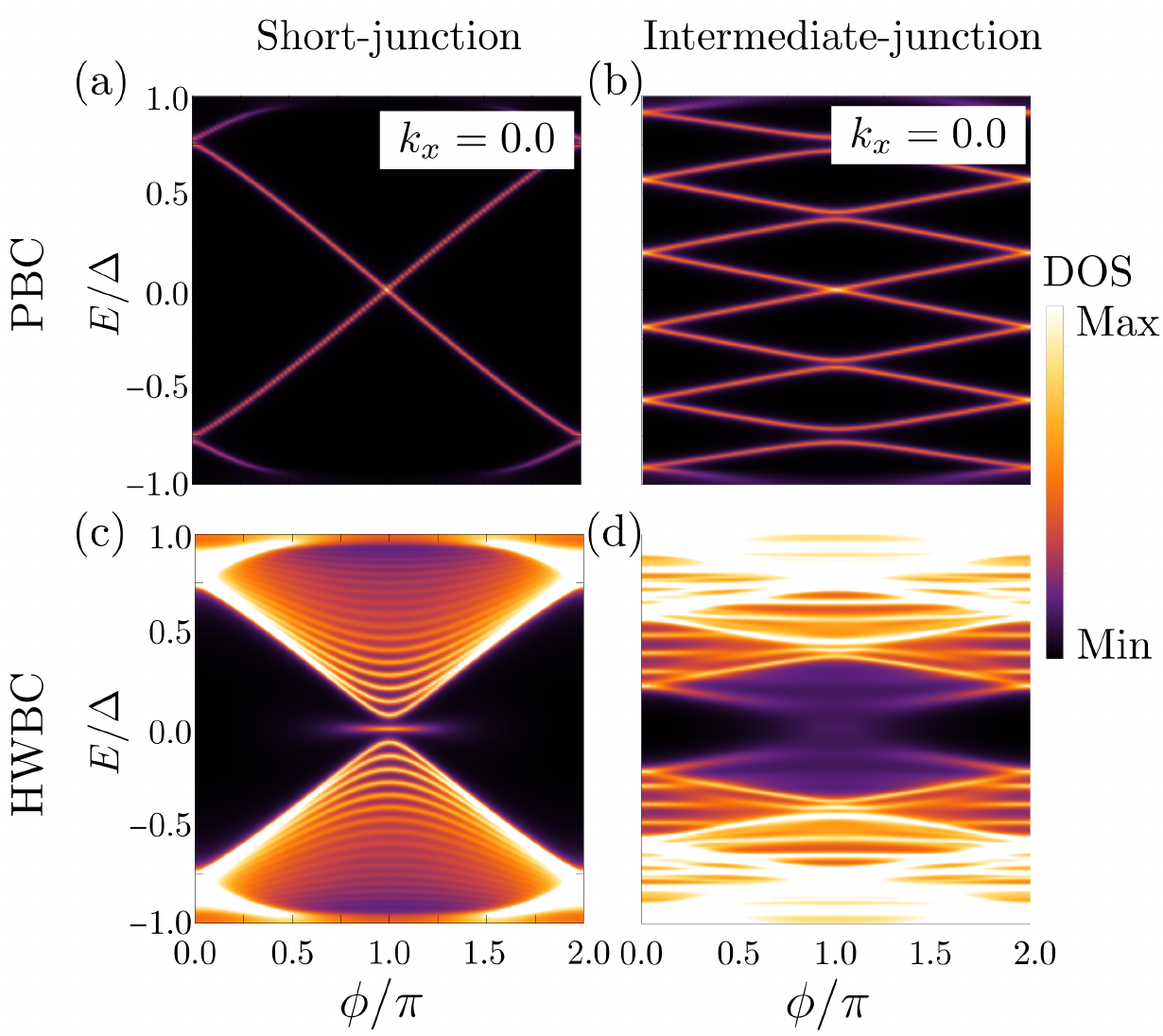}
    \caption{Low-doping regime ($\mu_\text{c}=0$): Phase-dependent DOS for a JJ with ${\cal Z}=0$, $\mu_\text{c} = 0$, $\Delta=0.8$, $\mu_s=0.8$ and (a, c) $n_y=2a$ and (b, d) $n_y=20a$ and PBC (a, b) and HWBC (c, d). 
     }
    \label{fig:ABS-LDOS}
\end{figure}

While the Chern number uniquely characterizes the bulk topology, the phase diagram of Fig.~\ref{fig:phase diagram}, reveals that distinct regions within the same topological sector can differ qualitatively in momentum space. This is illustrated by representative energy spectra shown in panels (a-h), each corresponding to a labeled region of the phase diagram. We see an example of this in regions (b), (c) and (h), with $\mathcal{C}=1$. Although they all exhibit the same Chern number, they differ on the $k_y$-point at which the chiral mode crosses zero energy: While regions (b) and (h) exhibit a crossing at $k_{y}=0$, (e), shows a zero energy crossing at $k_{y}=\pi/a$.  This difference will be important later when we discuss the transport results. Similarly, regions~(d)-(f) exhibit $\mathcal{C}=2$. Again, we observe two different scenarios: (d) and (f) exhibit zero energy crossings at arbitrary values of $k_y$, while (e) at high symmetry points $k_y=0$ and $k_y=\pi/a$.

In the following section, we focus on the zero Zeeman field case and analyze how the bulk topology with ${\cal C} = 1$ manifest in a JJ geometry.

\section{Josephson Junction at Zero Zeeman Field}\label{Sec.JJZero}

Having established the bulk topological phase diagram, we now turn to the analysis of the Josephson junction geometry shown in Fig.~\ref{fig:set-up}. In this section, we focus on the topological phase with $\mathcal{C}=1$ realized at zero Zeeman field ($\mathcal{Z}=0$) and chemical potentials $\mu_s \in (-2.5,2)$. We investigate the resulting Andreev bound states, local density of states, and transport properties, highlighting how bulk topology manifests in the junction spectra.

\begin{figure}[tb]
    \centering
    \includegraphics[width=1.0\linewidth]{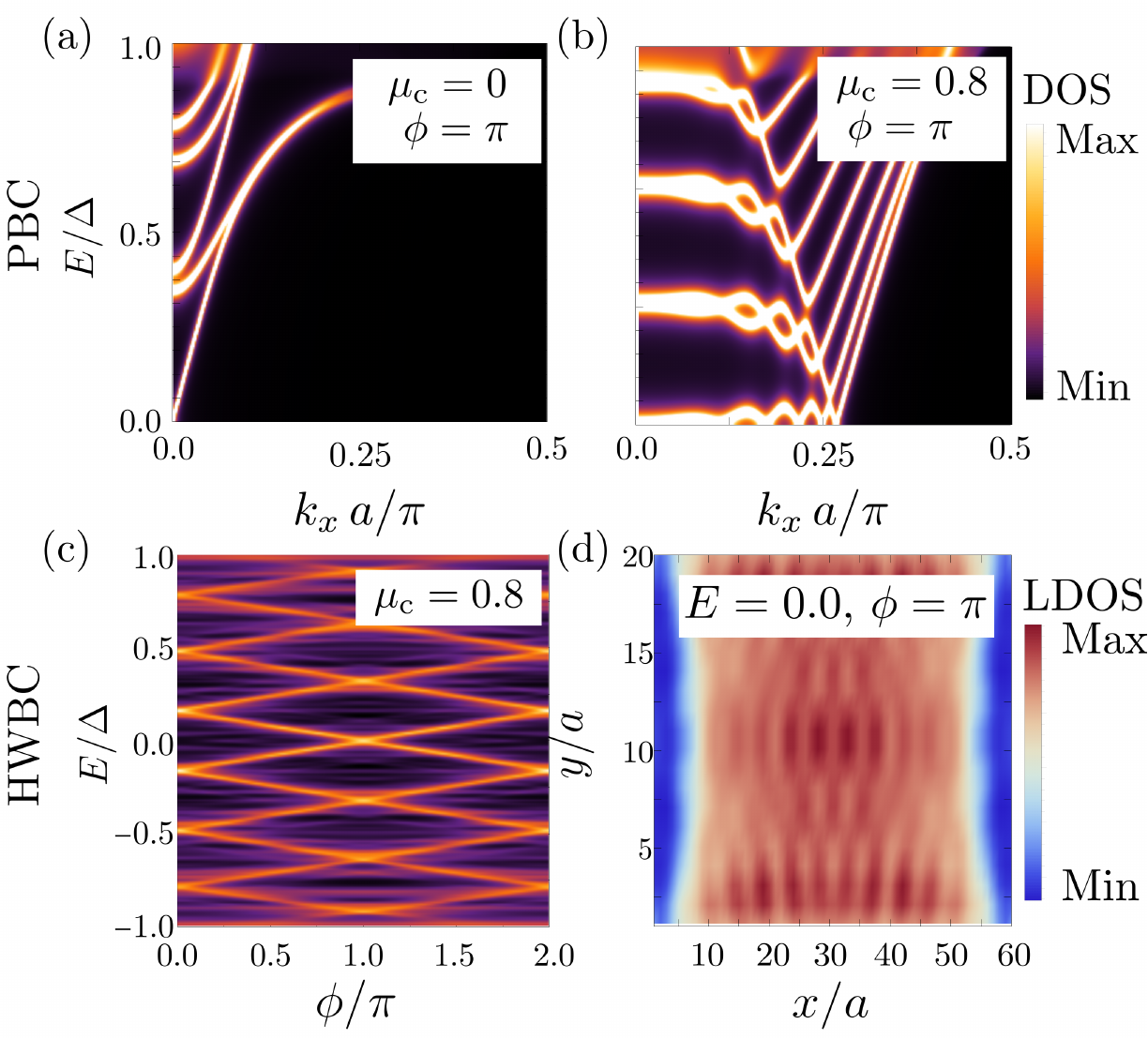}
    \caption{Finite-doping regime ($\mu_\text{c}=0.8$ and $n_y=20a$): (a, b) PBC as a function of $E$ and $k_x$, with (a) $\mu_\text{c}=0$ and (b) $\mu_\text{c}=0.8$ and $\phi=\pi$. (c, d) HWBC
    DOS as a function of energy and $\phi$ for the intermediate-junction regime, with $\mu_s=0.8$ and $\Delta=0.35$. The rest of the parameters are the same as in Fig.~\ref{fig:ABS-LDOS}. (d) LDOS as a function of $x$ and $y$ for $E=0$ and $\phi=\pi$. 
    }
    \label{fig:differentmodes} 
\end{figure}

\subsection{Andreev Bound States}
We begin by analyzing the Andreev bound states (ABS) of a Josephson junction in the short- and intermediate-junction regimes, characterized by $E_T/\Delta \sim 1$ and $E_T/\Delta \sim 0.1$, respectively. We find that the ABS spectra depend sensitively on the chemical potential of the central region, $\mu_\text{c}$. Accordingly, we distinguish between a low-doping regime, $\mu_\text{c}=0$, where the Fermi energy lies at the Dirac point $(k_x,k_y)=(0,0)$, and a finite-doping regime, $\mu_\text{c}\neq 0$, where the Fermi energy occurs at finite momentum.

\subsubsection{Zero-doping regime, $\mu_\text{c}=0$}
Figure~\ref{fig:ABS-LDOS} shows the ABS spectra in the low-doping regime for two types of boundary conditions. We consider periodic boundary conditions (PBC), which preserve translational invariance along the $x$-direction and allow for a good quantum number $k_x$, as well as hard-wall boundary conditions (HWBC), corresponding to a finite width $n_x=60a$ along $x$-direction.

The results with PBC $(k_x=0)$ exhibit the well-known energy-phase relation of other topological JJs, e.g.~JJ based on 2DTIs~\cite{Fu2009a} or 3DTI~\cite{Fu2008a}, with energy dispersion $E(\phi)=\Delta \cos(\phi/2)$, in the short-junction limit (a) and  $E(\phi)=E_T(\pi-|\phi|)/2$, in the long junction limit~(b), with the Thouless energy $E_T=\hbar v_\text{F}/n_y$. Interestingly, when we impose HWBC [Fig.~\ref{fig:ABS-LDOS}(c,d)], the ABS opens a gap at $\phi=\pi$ in both regimes, with a bound state at zero energy in the case of the short junction regime and a blurrier density profile in the intermediate regime. 

\begin{figure}[tb]
    \centering
    \includegraphics[width=1.0
    \linewidth]{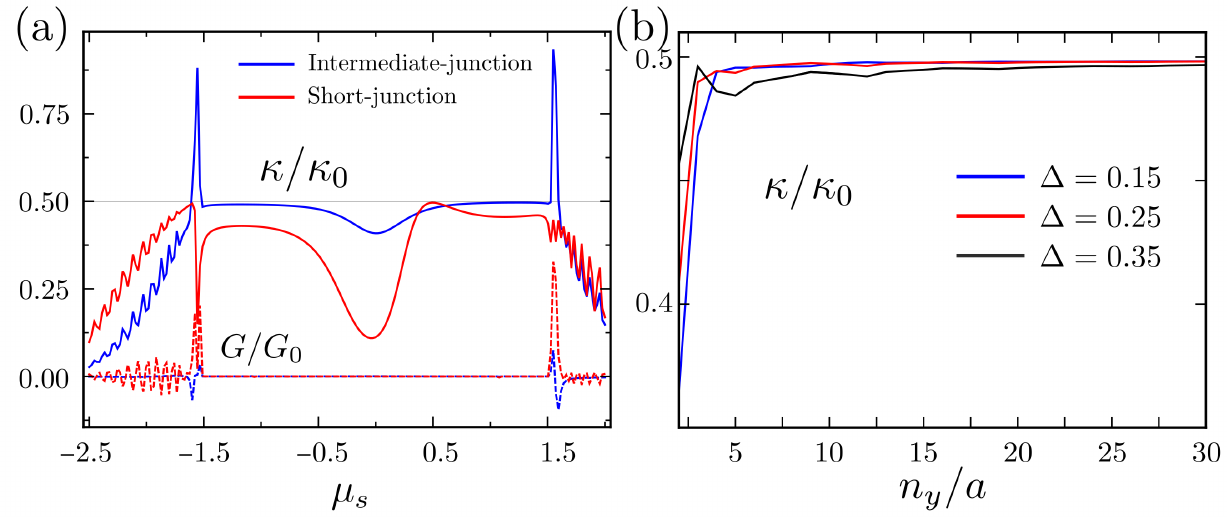}
    \caption{(a) Electrical (dashed) and thermal (solid) conductance as a function of $\mu_s$ for $\phi=\pi$ $\mu_\text{c}=0$, $\Delta=0.35$ and $\mathcal{Z}=0$, in the short-junction (red) and intermediate-junction (blue) regimes, with $n_y=2a$ and $n_y=20a$, respectively. (b) Thermal conductance as a function of the distance between superconducting leads $n_y$ for three different values of $\Delta$ and $E=0$, $\phi=\pi$, $\mu_s=1.0$ and $\mu_\text{c}=0.0$.}
    \label{fig:comparison}
\end{figure}

\subsubsection{Finite-doping regime, $\mu_\text{c}\neq 0$}
As the chemical potential of the central part is increased, more bands of the central region participate in the formation of ABS. Using PBC [Fig.~\ref{fig:differentmodes}(a,b)], we observe that for $\mu_\text{c}=0$, there is only one mode placed at $E=0$ and $k_x=0$, whereas for $\mu_\text{c}=0.8$, additional zero energy contributions appear at finite $k_x$. Each of these zero energy contributions can propagate along the $x$-direction introducing additional transport mechanisms with Andreev retroreflection and normal transmission, and therefore, we expect a change in both of electrical and thermal conductance in this regime~\cite{Titov2007a}. The ABS with HWBC becomes gapless at $\phi=\pi$ due to the participation of additional contributions at finite momentum, see Fig.~\ref{fig:differentmodes}(c), becoming similar to the PBC case, see Fig.~\ref{fig:ABS-LDOS}(b). Furthermore, the LDOS at $E=0$ and $\phi=\pi$ exhibits a fringe-like pattern resulting from the contribution of additional transverse subbands, see panel~(d).

\subsection{Electrical and Thermal Conductance}\label{sec. Transport}

We are now in position to evaluate the non-local electrical and thermal conductance in the linear-response regime of the zero energy state observed at $\phi=\pi$. Thus, we connect normal leads to the Josephson junction as sketched in Fig.~\ref{fig:set-up} and explore different doping for short- and intermediate-junction regimes. 

\subsubsection{Zero-doping regime, $\mu_\text{c}=0$}
In this regime, particles entering the junction from a normal lead perform specular Andreev reflections~\cite{Beenakker2006a} between the superconducting leads until they reach the opposite normal lead~\cite{Titov2007a}.
As a result, both electrical and thermal conductance acquire quantized values: The former is zero due to the neutral charge of the ABS, while the latter is half-quantized to $\kappa=0.5\kappa_0$, resulting from scattering with an excitation with equal superposition of electrons and holes. Crucially, the transmission along the junction depends on the presence of propagating modes connecting both leads. As we have seen in Fig.~\ref{fig:bulkbands}(c,d), the confinement in the $y$-direction, $n_y$ opens a gap at the Dirac cones that scales with the Thouless energy as $E_\text{conf}\propto (n+1/2) E_T$, with $E_T=\hbar v_\text{F}/n_y$. In this scenario, the presence of propagating modes in the $x$-direction occurs for $E_T\ll \Delta$, that is, as the junction is tuned into the intermediate- to long-junction regime~\cite{Titov2007a}.

To support the picture that a quantized response arises as we approach the long-junction regime, we compare the electrical and thermal conductance as a function of $\mu_s$ for both the short-junction and intermediate-junction regimes, see Fig.~\ref{fig:comparison}(a). We observe that the electrical conductance is suppressed for all values of $\mu_s$, except for $|\mu_s|=\sqrt{4m_0^2-\Delta^2}\approx 1.5$, where the superconducting gap closes and reopens, see Eq.~\eqref{eq:topsector2}. 
Moreover, the thermal conductance exhibits a similar behavior in both regimes, with a dip for $|\mu_s|<\Delta$ and an approximately flat profile for $\Delta\gtrsim |\mu_s|\geq \sqrt{4m_0^2-\Delta^2}$. For larger chemical potential $|\mu_s|>\sqrt{4m_0^2-\Delta^2}$, the conductance is suppressed. 
The main difference between both curves is that the intermediate-junction regime exhibits half-quantized values $\kappa/\kappa_0=0.5$ for $\Delta \lesssim |\mu_s| \leq \sqrt{4m_0^2-\Delta^2}$. 

\begin{figure}[tb]
    \centering
    \includegraphics[width=1.0\linewidth]{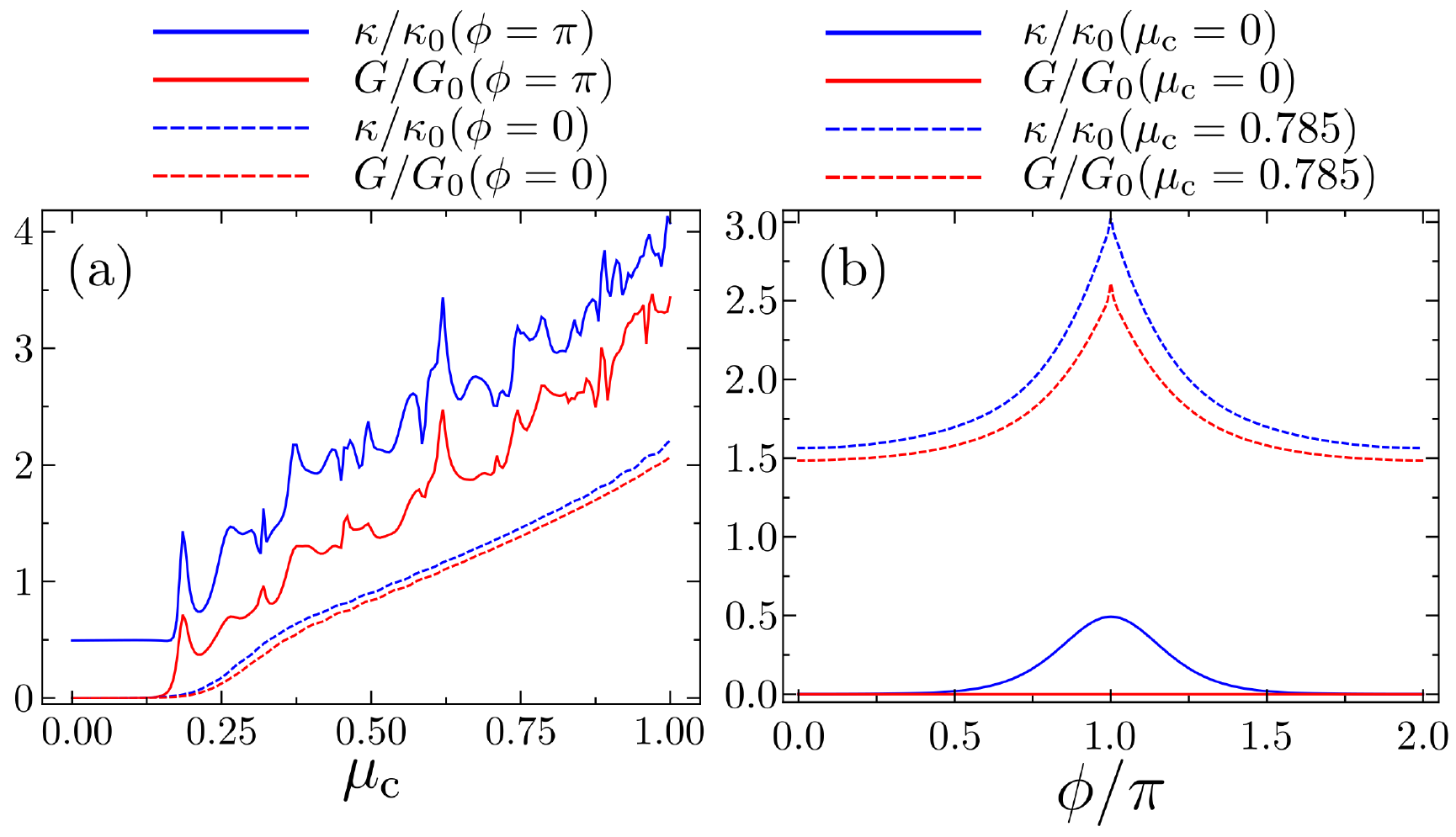}
    \caption{(a) Thermal (blue) and electrical (red) conductance for $\phi=\pi$ (solid line) and $\phi = 0$ (dashed line) for intermediate junction regime. (b) Thermal (blue) and electrical (red) conductance normalized by their respective quanta as function the superconducting phase difference for $\mu_\text{c}=0$ (solid line) and $\mu_\text{c} = 0.785$ (dashed lines).}
    \label{fig:transport-phase}
\end{figure}

In addition, in panel (b) we show the thermal conductance as a function of $n_y$ for three different values of $\Delta=0.15,\,0.25,\,0.35$, at $E=0$ and $\phi=\pi$. We see that as $n_y$ becomes larger than the coherence length ($\xi\sim 7a,\,4a,\, 3a$, respectively), the thermal conductance approaches the half-quantized value.

Interestingly, for $|\mu_s|>\sqrt{4m_0^2-\Delta^2}$, the thermal conductance becomes suppressed even though it is still in the topological phase, with $\mathcal{C}=1$. The reason for this is that the chiral Majorana modes cross zero energy at finite momentum, whereas the central part is essentially at zero momentum. Therefore, any coupling between the normal and superconducting sides is strongly suppressed since the normal part is gapped (by $m_0\neq 0$) at these points. 
We recover a full quantization setting $m_0=0$ also in the central part of the junction and normal leads, see further details in App.~\ref{App.comparison}.

\subsubsection{Finite-doping regime, $\mu_\text{c}\neq 0$} 
When the chemical potential of the normal region is increased, additional transport channels participate in the coupling of the left and right normal leads. Apart from increasing the transmission, these new transport channels allow for different transport mechanisms: While in the low-doping regime only specular electron-hole reflection is possible, at finite-doping, normal transmission and electron-hole retroreflection become active~\cite{Titov2007a}.
Thus, the combination of all processes results in an enhancement of both thermal and electrical conductance. Fig.~\ref{fig:transport-phase}(a) (dashed curves) shows that once the band of the central region begins to fill, the thermal conductance loses its quantization and the electrical conductance becomes finite. This transition is not a change of topology of the superconducting lead (the Chern number remains ${\cal C} = 1$), but a central region band-filling effect. As soon as additional transverse modes approach zero energy ($\mu_\text{c}\approx 3E_T$), they hybridize with the Majorana branch and destroy the single channel limit, producing the fringe-like patterns observed in the LDOS of Fig.~\ref{fig:ABS-LDOS}(c, d) and a non-quantized conductance, see further details in App.~\ref{App.comparison}. 

This distinction between zero and finite-doping regimes becomes even clearer in Fig.~\ref{fig:transport-phase}(a) and~(b), which shows the value of electrical and thermal conductance at  $\phi = 0$ and $\phi = \pi$ as a function of $\mu_\text{c}$. Here, we observe that a broad plateau persists around $\mu_\text{c}  = 0$, where the thermal conductance remains pinned to $\kappa = 0.5 \kappa_0$ and the electrical conductance vanishes. Around $\mu_\text{c} \approx 0.2$, the plateau ends abruptly as the normal region enters the conduction band and additional channels are activated. However, we predict quantized thermal conductance even if $\mu_\text{c}$ is finite, i.e.~beyond the Dirac point. This energy window is set by the energy difference between the bands in the central part, which is proportional to $E_T$ see Fig.~\ref{fig:bulkbands}(c,d) and App.~\ref{App.comparison}. 

\begin{figure}[tb]
    \centering
    \includegraphics[width=1.0\linewidth]{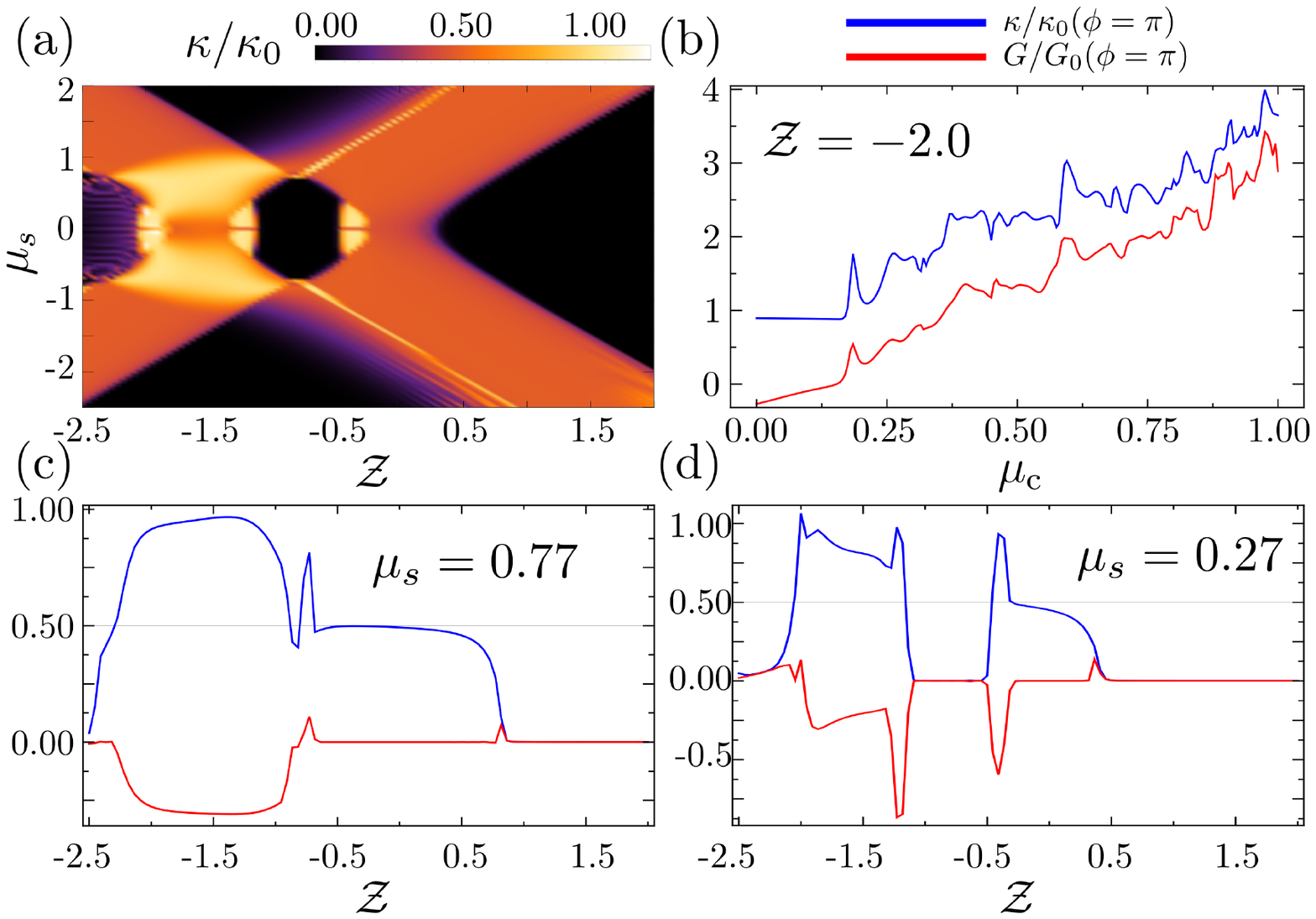}
    \caption{(a) Thermal conductance as a function of $\mathcal{Z}$ and $\mu_s$ for $\mu_\text{c}=0$ and $\delta=0.35$. (b) Thermal and electrical conductance as a function of $\mu_\text{c}$, for $\mu_s=0.8$, and $\Delta=0.35$ and $\mathcal{Z}=-2.0$. (c) and~(d) electrical and thermal conductance as a function of $\mathcal{Z}$ for $\mu_s=0.77$ and $\mu_s=0.27$, respectively. The rest of the parameters are the same as in panel (a).}
    \label{fig:Transport-Zeeman}
\end{figure}

\subsubsection{Phase difference dependence}
To complete our analysis, we now study the electrical and thermal conductance as a function of $\phi$, see Fig.~\ref{fig:transport-phase}(b). When the band of the central region is close to the Dirac point, the electrical conductance remains zero independently of $\phi$. In contrast, the thermal conductance exhibits a gaussian-like shape centered at $\phi = \pi$, reaching $\kappa/\kappa_0=0.5$ at $\phi=\pi$. For finite-doping both electrical and thermal conductances are finite and larger than 0.5, revealing the participation of channels at finite momentum. Also the shape changes from a gaussian-like shape to a decaying exponential one. A similar behavior has been observed in a proposal consisting on a JJ based on a proximitized three dimensional insulator, see Ref.~\onlinecite{Bauer2021a}. 
In contrast to our results, there, the thermal conductance quantization occurs already in the short-junction regime. 

\section{Josephson Junction at Finite Zeeman Field}\label{Sec.JJFinite}

We now analyze the four-terminal SNS junction in the presence of a finite Zeeman field (${\cal Z}\neq 0$), focusing on the intermediate-junction regime. A finite Zeeman field enlarges the accessible parameter space and stabilizes extended regions with Chern numbers ${\cal C}=1$ and ${\cal C}=2$ in the superconducting leads, as illustrated in the topological phase diagram of the top panel of Fig.~\ref{fig:phase diagram}. It is important to stress that this topological classification refers to an isolated superconducting lead, as discussed in Sec.~\ref{Sec.JJZero}.

Figure~\ref{fig:Transport-Zeeman}(a) presents a color plot with the thermal conductance of the SNS junction as a function of the Zeeman field ${\cal Z}$ and the superconducting chemical potential $\mu_s$ in the zero-doping regime $\mu_\text{c}=0$. The global structure closely reflects the phase diagram of the isolated lead: trivial regions (${\cal C}=0$) display a strongly suppressed thermal response. By contrast, phases with ${\cal C}=1$ [regions (b) and (h) in Fig.~\ref{fig:phase diagram}] give rise to extended plateaus near $\kappa = 0.5\kappa_0$, whereas region (c) in the top panel of Fig.~\ref{fig:phase diagram}, despite sharing the same Chern number, does not exhibit quantization. The correspondence becomes even less direct in the ${\cal C}=2$ phase, where the thermal conductance is generally reduced and reaches its maximal value only in restricted regions of parameter space. We return to this point below.

\begin{figure}[tb]
    \centering
    \includegraphics[width=1.0\linewidth]{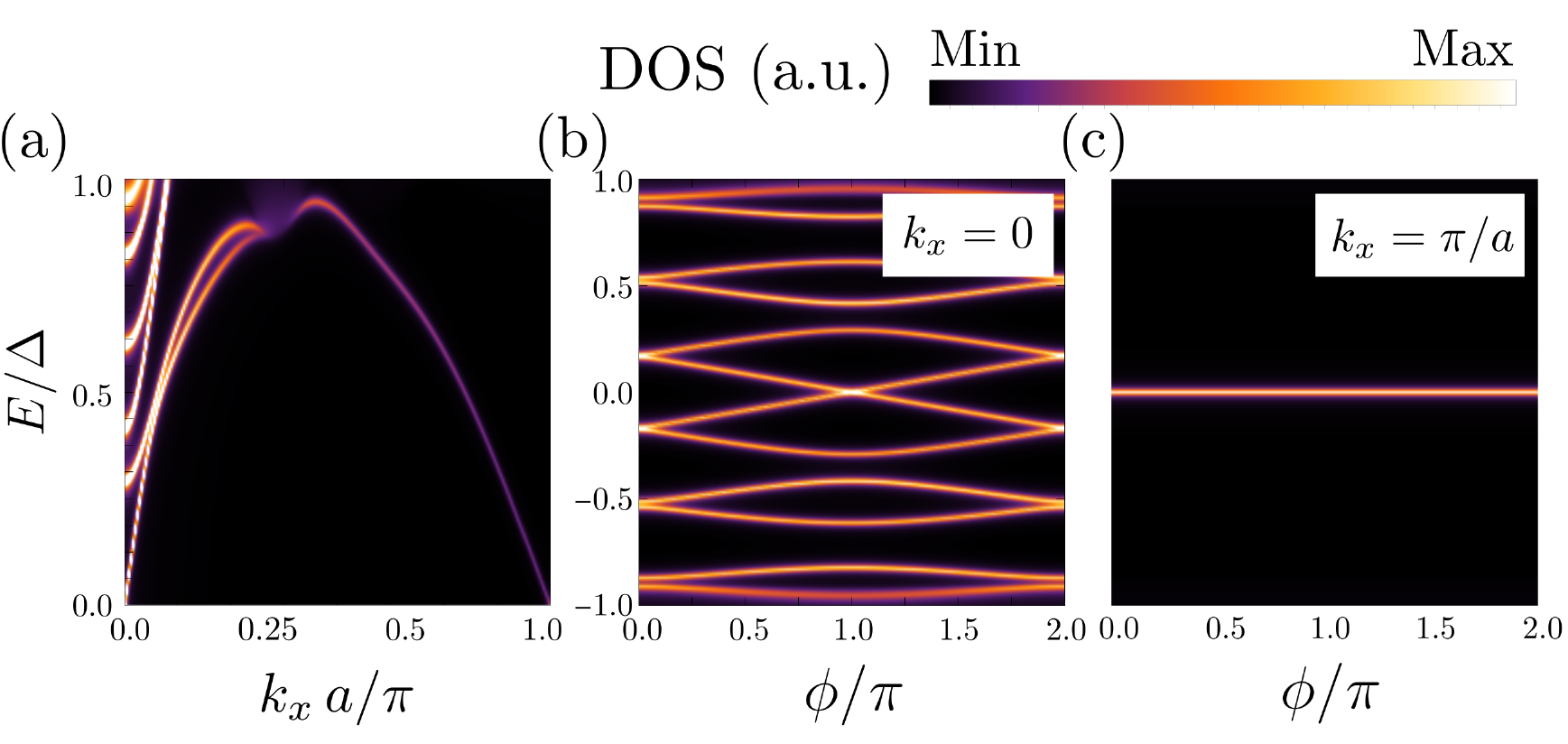}
    \caption{(a) DOS as a function of $k_x$ and $E$ for a Josephson junction with PBC, $\phi=\pi$. The rest of parameters are set to $n_y=20 a$, $m_0=0.8$, $\mu_s=1.0$, $\Delta=0.35$, $\mathcal{Z}=-1.0$ and $\mu_\text{c}=0$. (b), (c) DOS as a function of $\phi$ for $k_x=0$ and $k_x=\pi/a$ The rest of parameters are the same as in panel (a).}
    \label{fig:C2}
\end{figure}

In Fig.~\ref{fig:Transport-Zeeman}(c) and (d) we show line cuts of the electrical and thermal conductance as functions of $\mathcal{Z}$ for fixed values of $\mu_s=0.77$ and $\mu_s=0.27$, respectively. For $\mu_s=0.77$ [panel (c)], a broad interval within the $\mathcal{C}=1$ phase exhibits half-quantized thermal conductance accompanied by vanishing electrical conductance, consistent with Fig.~\ref{fig:comparison}. In addition, for $-2.2\lesssim \mathcal{Z}\lesssim -0.7$ the thermal conductance approaches $\kappa/\kappa_0\approx 1$, now accompanied by a finite electrical conductance. A qualitatively different behavior is observed for $\mu_s=0.27$ [panel (d)], where no clear quantization emerges. This is consistent with the absence of quantized plateaus in Fig.~\ref{fig:comparison} for $\mu_s\lesssim \Delta$. Panel (b) further shows the electrical and thermal conductance as functions of the central chemical potential $\mu_\text{c}$, revealing a behavior similar to the $\mathcal{Z}=0$ case in the $\mathcal{C}=1$ phase. In contrast to that situation, the thermal conductance reaches $\kappa/\kappa_0\approx 1$ and the electrical conductance becomes finite, reflecting the fact that two chiral Majorana modes combine into an ordinary fermionic channel.

The role of the normal region is isolated in Fig.~\ref{fig:Transport-Zeeman}(b), where ${\cal Z}$ and $\mu_s$ are fixed while the chemical potential in the central region is varied. The persistence of the thermal plateau up to the onset of bulk-band filling indicates that the reduced thermal conductance in the ${\cal C}=2$ phase is not primarily controlled by normal-region doping, but instead originates from the momentum-space structure of the Majorana modes inherited from the superconducting leads.

Further insight into the ${\cal C}=2$ regime is obtained by examining the ABS under periodic boundary conditions, shown in Fig.~\ref{fig:C2}. Panel (a) displays the ABS as functions of energy and $k_x$ at $\phi=\pi$, revealing two distinct zero-energy states located at $k_x=0$ and $k_x=\pi/a$. Beyond their different spectral weights, these states exhibit markedly different phase dependencies, as shown in panels (c) and (d). The mode at $k_x=0$ disperses with the superconducting phase, while the mode at $k_x=\pi/a$ remains essentially phase independent, indicating that it does not originate from coherent coupling between the two superconducting leads. Consequently, its contribution to nonlocal transport is expected to be strongly suppressed. This behavior contrasts with other NSN junctions hosting topological superconductivity, where the thermal conductance reproduces the topological phase diagram~\cite{Ning-Xuan2022a}.

\section{Conclusions}\label{Sec.Conclusions}

In this work, we have investigated the interplay between topology and junction geometry in the transport properties of a four-terminal Josephson junction with transverse chiral topological superconducting leads. By combining a lattice-regularized Dirac-BdG model with a non-equilibrium Green's function approach, we have established concrete conditions under which bulk topological properties of the superconducting leads manifest in experimentally accessible nonlocal electrical and thermal conductances.

A central result of our study is that combined thermal and electrical transport provides a clear signature of chiral Majorana modes. 
In the ${\cal C}=1$ phase, a single chiral Majorana channel gives rise to a robust half-quantized thermal conductance at a superconducting phase difference $\phi=\pi$, while the nonlocal electrical conductance remains strongly suppressed due to particle-hole symmetry. Importantly, this behavior persists at finite doping in the intermediate- to long-junction regimes, where a single propagating mode dominates transport through the normal region.

At finite Zeeman fields, we found that the thermal conductance continues to reflect the topology of the isolated superconducting leads in the ${\cal C}=1$ phase, yielding extended half-quantized plateaus. In contrast, in the ${\cal C}=2$ phase, the thermal response is no longer uniquely fixed by the Chern number: depending on the momentum-space location of the Majorana modes, the thermal conductance can be either strongly suppressed or approach $\kappa\simeq\kappa_0$.
This result provides a natural explanation for why thermal transport measurements may fail to directly reproduce bulk phase diagrams in multiterminal geometries.

These findings highlight an important distinction between bulk topology and transport in finite multiterminal geometries. While the Chern number determines the number of chiral Majorana modes supported by an isolated superconducting lead, the efficiency with which these modes contribute to heat transport across a junction depends sensitively on their momentum-space structure, spatial localization, and mutual interference. 
As a result, the junction geometry should be regarded as a selective probe of Majorana modes rather than a direct measure of the bulk topological invariant alone.

More broadly, our results demonstrate that interpreting transport signatures of topological superconductivity requires careful consideration of geometry, band filling, and mode hybridization. The framework developed here provides practical guidance for designing thermal-transport experiments aimed at identifying chiral Majorana modes, and can be readily extended to other multiterminal 
platforms, including helical topological superconductors, and higher-order topological insulators. It thus provides a systematic route to extract topological signatures from thermal conductance measurements in Majorana-based devices.

\begin{acknowledgments}
We acknowledge stimulating discussions with M. P. Stehno and S. U. Piatrusha. We acknowledge funding support from the Deutsche Forschungsgemeinschaft (DFG, German Research Foundation) under Germany’s Excellence Strategy through the W\"urzburg-Dresden Cluster of Excellence on Complexity and Topology in Quantum Matter ct.qmat (EXC 2147, Project ID
390858490) as well as through the Collaborative Research
Center SFB 1170 ToCoTronics (Project ID 258499086).
We also gratefully acknowledge the Gauss Centre for Supercomputing e.V. (\url{www.gauss-centre.eu}) for funding this project by providing computing time on the GCS Supercomputer SuperMUC-NG at Leibniz Supercomputing Center (\url{www.lrz.de}).
\end{acknowledgments}

\begin{figure}[tb]
    \centering
    \includegraphics[width=0.92\linewidth]{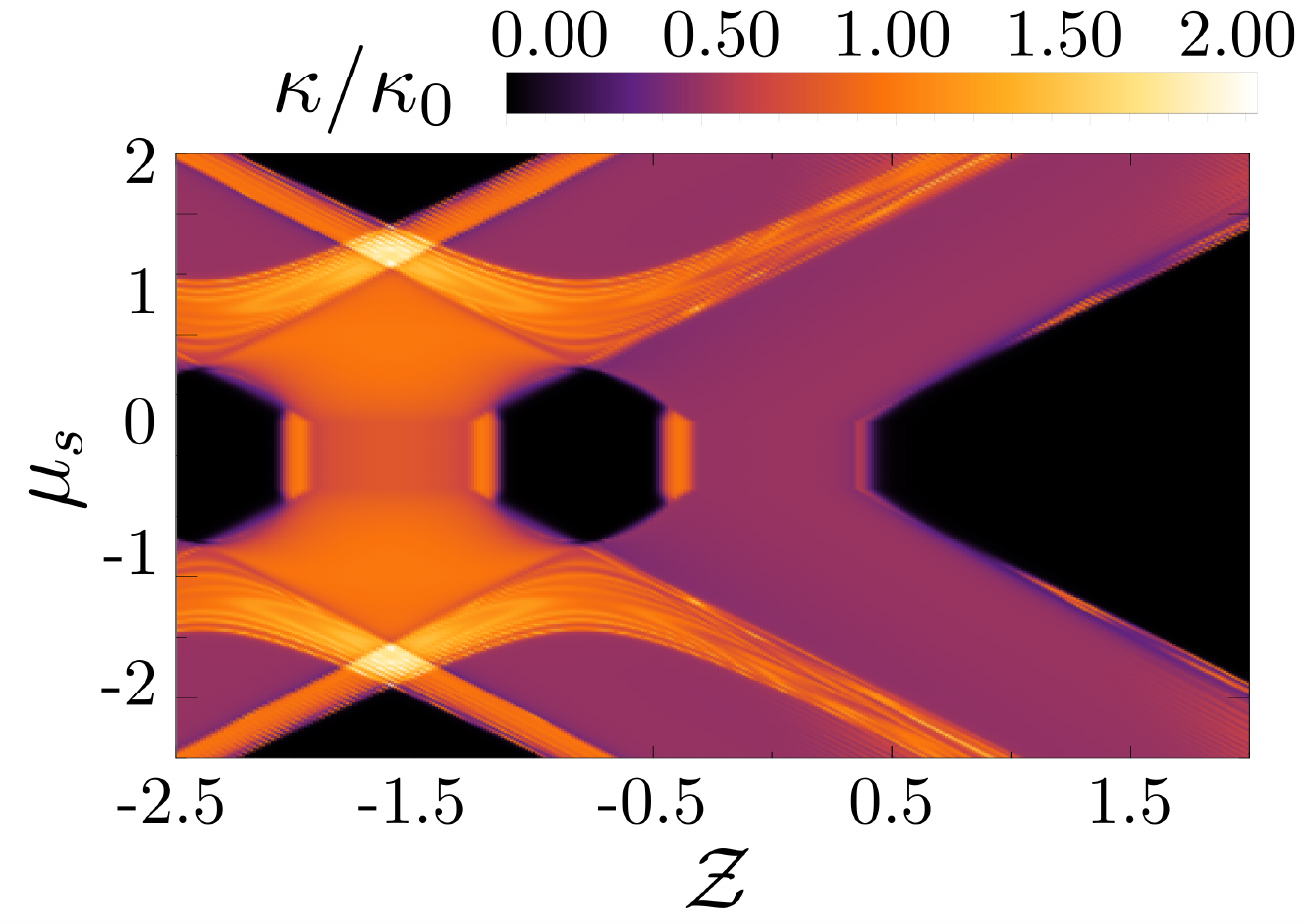}
    \caption{Thermal conductance as a function of $\mathcal{Z}$ and $\mu_s$ for the intermediate-junction $(n_y=20a)$ regime, with $\mu_c=0.0$ and $\phi=\pi$. The Wilson mass is set to zero within the normal part (including the leads) and $m_0=0.8$ in the superconducting leads.}
    \label{fig:normalmass0}
\end{figure}

\appendix

\section{Green's function}\label{App.GF}

We evaluate the transport properties using Green's functions. Coupling the finite scattering region to four semi-infinite leads produces the retarded (advanced) Green's function of the central system,
${\cal G}^{r}_{il}\left({\cal G}^{a}_{li}\right)$, 
\begin{equation}
    \label{eq:Dyson}
    {\cal G}^{r} = \left[\varepsilon+i\eta-H_{\text{c}}-\Sigma^r(\phi)\right]^{-1},
\end{equation}
with $H_{\text{c}}$ the Hamiltonian of the normal central part and $\eta=10^{-5}$ the Dynes parameter that distinguishes between retarded $(\eta\to0^+)$ and advanced $(\eta\to0^-)$ Green's functions.  

We account for the coupling between the central scattering region and the leads by the addition of the self-energy $\Sigma^r(\phi)=\Sigma^r_{L}+\Sigma^r_{R}+\Sigma^r_{T}(\varphi_T)+\Sigma^r_{B}(\varphi_B)$ and 
$\Sigma_{i}^{r,a}(\varepsilon) = V_{\text{c},i} g_{{\rm surf},i}^{r,a} (\varepsilon )V_{i,\text{c}}$
with $V_{\text{c},i}$ the coupling matrix between the $i$-lead and the central part.
In practice, we compute the retarded surface Green's function of the $i$-lead $[g_{{\rm surf},i}^r(\varepsilon)]$ with an iterative decimation scheme~\cite{sancho1985highly}.
Finally, the coupling rates between the central Hamiltonian and the normal $k$-lead is $\Gamma_k =i\left(\Sigma_k^r-\Sigma_k^a\right)$.

\section{Thermal conductance quantization} \label{App.comparison}

In this section, we show additional details of the interplay of $m_0$, $\mu_\text{c}$ and the Zeeman field intensity $\mathcal{Z}$ in the thermal conductance quantization.

\begin{figure}[tb]
    \centering
    \includegraphics[width=1.0\linewidth]{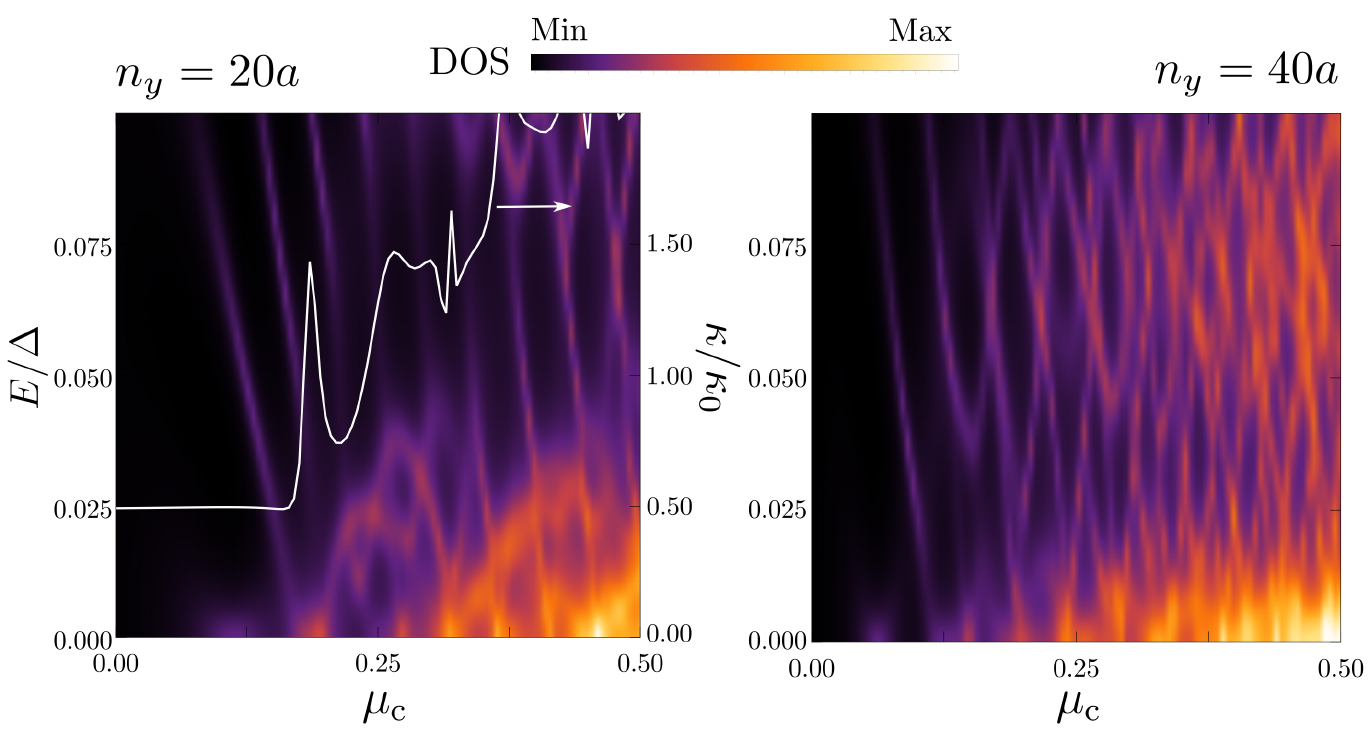}
    \caption{DOS at $\phi=\pi$ as a function of $\mu_\text{c}$ and $E$ for $\mu_s=0.785$ and $\Delta=0.35$, with $n_y=20a$ (left) and $n_y=40a$ (right). Moreover, on the left panel we have superposed the thermal conductance as a function of $\mu_\text{c}$ in white line.}
    \label{fig:gap_vs_muc}
\end{figure}

\subsection{Importance of the Wilson mass on the normal part}\label{app:mass}

In the main text, we have built the four-terminal junction using a constant Wilson mass $m_0=0.8$ for the whole system. As we have seen, the presence of this mass breaks time-reversal symmetry, opening a gap at the three Dirac cones located at finite momentum, i.e.~$(k_x,k_y)=(0,\pi/a),~(\pi/a,0)$ and $(\pi/a,\pi/a)$. Although the topological superconducting phase needs a finite mass, see Eqs.~\eqref{eq:topsector1} and~\eqref{eq:topsector2}, the normal regions do not. Indeed, some results can change if we set $m_0=0$ in the normal region, viz.~those that have shown crossings of a chiral Majorana mode at finite momentum. 
To see the changes introduced by this heterogeneous mass, we have recalculated Fig.~\ref{fig:Transport-Zeeman}(a), now with $m_0=0$ within the normal part, see Fig.~\ref{fig:normalmass0}. Here, we observe that the regions with $\mathcal{C}=1$ and with a crossing at $k_y=0$, exhibit no difference [(b) and (h) in Fig.~\ref{fig:phase diagram}]. In contrast, region (c), shows now a half-quantized region.
The rest of the phase diagram shows more resemblance with the phase diagram, but as expected, no quantization is observed, with thermal conductance values between 0 and 2$\kappa_0$, due to the additional contribution of the Dirac cones.

\subsection{Role of the central region doping $\mu_\text{c}$}\label{app:centraldoping}

\begin{figure}[tb]
    \centering
    \includegraphics[width=0.92\linewidth]{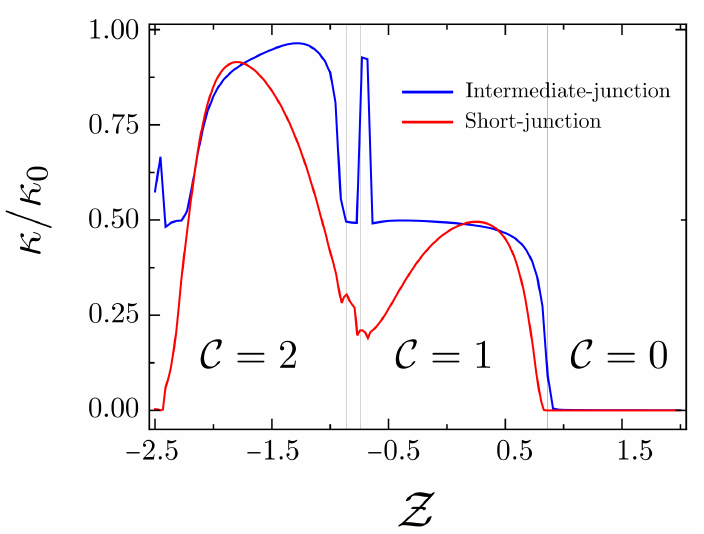}
    \caption{Thermal conductance as a function of $\mathcal{Z}$ for the short-junction $(n_y=2a)$ and intermediate-junction $(n_y=20a)$ regimes, with $\mu_s=0.785$ and $\mu_c=0.0$.
    Thermal conductance as a function of $\mathcal{Z}$ for the short-(red, $n_y=2a$) and intermediate-junction (blue, $n_y=20a$) regimes.
    While the intermediate junction exhibits a robust half-quantized plateau in the ${\cal C}=1$ phase, the short junction presents a suppressed quantization due to enhanced mode hybridization. Other parameters are $\mu_s=0.785$ and $\mu_c=0.0$.
    }
    \label{fig:comparisonzeeman}
\end{figure}

As we have seen in the main text, the occurrence of thermal conductance quantization $\kappa/\kappa_0=0.5$ depends on the doping of the central part. Indeed, the quantization extends over a window of approximately $\mu_\text{c}\approx 0.2$, see Fig.~\ref{fig:differentmodes}(a). This energy window is set by the level spacing imposed by the confinement $n_y$, i.e.~$E_\text{conf}\approx 1.5 E_T(n+1/2)$, with $n\in \mathbb{Z}$ and $E_T=\hbar v_\text{F}/n_y$. When the second level $(n=1)$ approaches zero energy ($\mu_\text{c}\approx E_1$), additional channels contribute to the electrical and thermal conductance, resulting in a lack of quantization. 
To show this effect, we compare the DOS at $\phi=\pi$ as a function of $\mu_\text{c}$ and $E$ for two Josephson junctions, with $n_y=20a$ (left) and $n_y=40a$ (right), see Fig.~\ref{fig:gap_vs_muc}. In the left panel, we observe that the presence of a finite DOS at zero energy $(\mu_\text{c}\approx 0.2)$ increases the quantization of thermal conductance (see the white curve superposed in the left panel). Thus, a Josephson junction with $n_y=40a$, further reduces the doping window where one can observe thermal conductance quantization to $0.125$, see right panel of Fig.~\ref{fig:gap_vs_muc}.

\subsection{Comparison of short-junction and intermediate junction regime}\label{app:short-intermediate}
In Fig.~\ref{fig:comparisonzeeman}, we show a comparison between the intermediate and short-junction regimes as a function of the Zeeman intensity. Under these conditions, the superconductor goes through different topological phases, i.e.:~$\mathcal{C}=0,\,1,\,2$. We see as in Fig.~\ref{fig:comparisonzeeman} that the thermal conductance is half-quantized within the $\mathcal{C}=1$ phase. In contrast, the short-junction regime shows an absence of quantization. For the $\mathcal{C}=2$ phase, we observe an absence of quantization in both the intermediate- and short-junction regimes. 

In Fig.~\ref{fig:comparisonzeeman}, we compare the thermal conductance in the short- and intermediate-junction regimes as a function of  the Zeeman field ${\cal Z}$. As this parameter is varied, the superconducting leads tranverse distinct topological phases characterized by Chern numbers ${\cal C} =0,\,1$ and $2$.

In the intermediate-junction regime, the thermal conductance exhibits a clear half-quantized plateau within the ${\cal C}=1$ phase. By contrast, in the short-junction regime the quantization is strongly suppressed. This difference originates from the role of the normal region in mediating transport. In the intermediate-junction limit ($E_T\ll \Delta$), transport through the junction is dominated by a single propagating mode in the normal region, which couples selectively to the chiral Majorana mode in the superconducting leads. As a result, heat transport is effectively carried by a single channel, yielding $\kappa \simeq 0.5\, \kappa_0$.

In the short-junction regime ($E_T\sim \Delta$), however, the two superconducting leads hybridize more strongly across the normal region. Multiple transverse modes contribute to transport, and continuum states near the gap edges interfere with the Majorana channel. 
This additional mixing-mode causes a non-universal transmission probability preventing roboust quantization.

Upon entering the ${\cal C}=2$ phase, quantization is absent in both regimes. In this case, the presence of two chiral Majorana modes does not automatically lead to $\kappa=\kappa_0$, since their momentum-space structure and coupling to the normal region determine whether they contribute coherently to heat transport.
Therefore, the intermediate-junction regime acts as a momentum and mode filter that isolates the Majorana contribution, while the short junction allows competing channels that spoil quantization.

\bibliography{refs}

@article{qi2010chiral,
  title={Chiral topological superconductor from the quantum Hall state},
  author={Qi, Xiao-Liang and Hughes, Taylor L and Zhang, Shou-Cheng},
  journal={Phys. Rev. B },
  volume={82},
  number={18},
  pages={184516},
  year={2010},
  publisher={APS}
}

@article{caroli1971direct,
  title={Direct calculation of the tunneling current},
  author={Caroli, CNRPC and Combescot, Ro and Nozieres, Ph and Saint-James, D},
  journal={Journal of Physics C: Solid State Physics},
  volume={4},
  number={8},
  pages={916},
  year={1971},
  publisher={IOP Publishing}
}

@article{chiu2016classification,
  title={Classification of topological quantum matter with symmetries},
  author={Chiu, Ching-Kai and Teo, Jeffrey CY and Schnyder, Andreas P and Ryu, Shinsei},
  journal={Reviews of Modern Physics},
  volume={88},
  number={3},
  pages={035005},
  year={2016},
  publisher={APS}
}

@article{Arrachea2025a,
  title={Thermoelectric Processes of Quantum Normal-Superconductor Interfaces},
  author={Arrachea, Liliana and Braggio, Alessandro and Burset, Pablo and Lee, Eduardo JH and Levy Yeyati, Alfredo and S{\'a}nchez, Rafael},
  journal={Annalen der Physik},
  volume={537},
  number={11},
  pages={e00197},
  year={2025},
  publisher={Wiley Online Library}
}

@article{Mateos2024a,
  title={Nonlocal thermoelectricity in quantum wires as a signature of Bogoliubov-Fermi points},
  author={Mateos, Juan Herrera and Tosi, Leandro and Braggio, Alessandro and Taddei, Fabio and Arrachea, Liliana},
  journal={Physical Review B},
  volume={110},
  number={7},
  pages={075415},
  year={2024},
  publisher={APS}
}

@article{Ruiz2024a,
  title={Binding zero modes with fluxons in Josephson junctions of time-reversal invariant topological superconductors},
  author={Ruiz, Gabriel F Rodr{\'\i}guez and Reich, Adrian and Shnirman, Alexander and Schmalian, J{\"o}rg and Arrachea, Liliana},
  journal={Physical Review B},
  volume={110},
  number={24},
  pages={245427},
  year={2024},
  publisher={APS}
}

@article{sancho1985highly,
  title={Highly convergent schemes for the calculation of bulk and surface Green functions},
  author={Sancho, MP Lopez and Sancho, JM Lopez and Sancho, JM Lopez and Rubio, J},
  journal={Journal of physics. F, Metal physics},
  volume={15},
  number={4},
  pages={851--858},
  year={1985},
  publisher={Institute of Physics}
}

@article{Titov2007a,
  title = {Excitation gap of a graphene channel with superconducting boundaries},
  author = {Titov, M. and Ossipov, A. and Beenakker, C. W. J.},
  journal = {Phys. Rev. B},
  volume = {75},
  issue = {4},
  pages = {045417},
  numpages = {8},
  year = {2007},
  month = {Jan},
  publisher = {American Physical Society},
  doi = {10.1103/PhysRevB.75.045417},
  url = {https://link.aps.org/doi/10.1103/PhysRevB.75.045417}
}

@article{Bittermann2024a,
  title = {Photonic cross-noise spectroscopy of Majorana bound states},
  author = {Bittermann, L. and Dominguez, F. and Recher, P.},
  journal = {Phys. Rev. B},
  volume = {110},
  issue = {4},
  pages = {045429},
  numpages = {15},
  year = {2024},
  month = {Jul},
  publisher = {American Physical Society},
  doi = {10.1103/PhysRevB.110.045429},
  url = {https://link.aps.org/doi/10.1103/PhysRevB.110.045429}
}

@article{Dominguez2024a,
  title = {Fraunhofer pattern in the presence of Majorana zero modes},
  author = {Dominguez, F. and Novik, E. G. and Recher, P.},
  journal = {Phys. Rev. Res.},
  volume = {6},
  issue = {2},
  pages = {023304},
  numpages = {17},
  year = {2024},
  month = {Jun},
  publisher = {American Physical Society},
  doi = {10.1103/PhysRevResearch.6.023304},
  url = {https://link.aps.org/doi/10.1103/PhysRevResearch.6.023304}
}

@misc{traverso2025a,
      title={Confinement-induced Majorana modes in a nodal topological superconductor}, 
      author={Simone Traverso and Niccolò Traverso Ziani and Maura Sassetti and Fernando Dominguez},
      year={2025},
      eprint={2407.06925},
      archivePrefix={arXiv},
      primaryClass={cond-mat.mes-hall},
      url={https://arxiv.org/abs/2407.06925}, 
}

@misc{heinz2024a,
      title={Interedge backscattering in time-reversal symmetric quantum spin Hall Josephson junctions}, 
      author={Cajetan Heinz and Patrik Recher and Fernando Dominguez},
      year={2024},
      eprint={2410.12357},
      archivePrefix={arXiv},
      primaryClass={cond-mat.mes-hall},
      url={https://arxiv.org/abs/2410.12357}, 
}

@article{Finocchiaro2018a,
  title = {Topological $\pi$ Junctions from Crossed Andreev Reflection in the Quantum Hall Regime},
  author = {Finocchiaro, F. and Guinea, F. and San-Jose, P.},
  journal = {Phys. Rev. Lett.},
  volume = {120},
  issue = {11},
  pages = {116801},
  numpages = {6},
  year = {2018},
  month = {Mar},
  publisher = {American Physical Society},
  doi = {10.1103/PhysRevLett.120.116801},
  url = {https://link.aps.org/doi/10.1103/PhysRevLett.120.116801}
}

@article{Fleckenstein2021a,
  title = {Formation and detection of Majorana modes in quantum spin Hall trenches},
  author = {Fleckenstein, C. and Ziani, N. Traverso and Calzona, A. and Sassetti, M. and Trauzettel, B.},
  journal = {Phys. Rev. B},
  volume = {103},
  issue = {12},
  pages = {125303},
  numpages = {12},
  year = {2021},
  month = {Mar},
  publisher = {American Physical Society},
  doi = {10.1103/PhysRevB.103.125303},
  url = {https://link.aps.org/doi/10.1103/PhysRevB.103.125303}
}

@article{Prada2012a,
  title = {Transport spectroscopy of $NS$ nanowire junctions with Majorana fermions},
  author = {Prada, Elsa and San-Jose, Pablo and Aguado, Ram\'on},
  journal = {Phys. Rev. B},
  volume = {86},
  issue = {18},
  pages = {180503},
  numpages = {5},
  year = {2012},
  month = {Nov},
  publisher = {American Physical Society},
  doi = {10.1103/PhysRevB.86.180503},
  url = {https://link.aps.org/doi/10.1103/PhysRevB.86.180503}
}

@article{Fleckenstein2018a,
  title = {Decaying spectral oscillations in a Majorana wire with finite coherence length},
  author = {Fleckenstein, C. and Dom\'{\i}nguez, F. and Traverso Ziani, N. and Trauzettel, B.},
  journal = {Phys. Rev. B},
  volume = {97},
  issue = {15},
  pages = {155425},
  numpages = {10},
  year = {2018},
  month = {Apr},
  publisher = {American Physical Society},
  doi = {10.1103/PhysRevB.97.155425},
  url = {https://link.aps.org/doi/10.1103/PhysRevB.97.155425}
}

@article{Haim2015a,
  title = {Signatures of Majorana Zero Modes in Spin-Resolved Current Correlations},
  author = {Haim, Arbel and Berg, Erez and von Oppen, Felix and Oreg, Yuval},
  journal = {Phys. Rev. Lett.},
  volume = {114},
  issue = {16},
  pages = {166406},
  numpages = {6},
  year = {2015},
  month = {Apr},
  publisher = {American Physical Society},
  doi = {10.1103/PhysRevLett.114.166406},
  url = {https://link.aps.org/doi/10.1103/PhysRevLett.114.166406}
}

@article{Lutchyn2010a,
  title = {Majorana Fermions and a Topological Phase Transition in Semiconductor-Superconductor Heterostructures},
  author = {Lutchyn, Roman M. and Sau, Jay D. and Das Sarma, S.},
  journal = {Phys. Rev. Lett.},
  volume = {105},
  issue = {7},
  pages = {077001},
  numpages = {4},
  year = {2010},
  month = {Aug},
  publisher = {American Physical Society},
  doi = {10.1103/PhysRevLett.105.077001},
  url = {https://link.aps.org/doi/10.1103/PhysRevLett.105.077001}
}

@article{Oreg2010a,
  title = {Helical Liquids and Majorana Bound States in Quantum Wires},
  author = {Oreg, Yuval and Refael, Gil and von Oppen, Felix},
  journal = {Phys. Rev. Lett.},
  volume = {105},
  issue = {17},
  pages = {177002},
  numpages = {4},
  year = {2010},
  month = {Oct},
  publisher = {American Physical Society},
  doi = {10.1103/PhysRevLett.105.177002},
  url = {https://link.aps.org/doi/10.1103/PhysRevLett.105.177002}
}

@article{Alicea2012a,
   author    = {Jason Alicea},
   title     = {New directions in the pursuit of Majorana fermions in solid state systems},
   journal   = {Rep. Prog. Phys.},
   year      = {2012},
   volume    = {75},
   pages     = {076501}
}

@article{Das2012a,
   author    = {Das, Anindya and Ronen, Yuval and Most, Yonatan and Oreg, Yuval and Heiblum, Moty and Shtrikman, Hadas},
   title     = {Zero-bias peaks and splitting in an Al-InAs nanowire topological superconductor as a signature of Majorana fermions},
   journal   = {Nat. Phys.},
   year      = {2012},
   volume    = {8},
   doi       = {10.1038/nphys2479},
   pages     = {887}
}

@article{Bocquillon2016a,
  title={Gapless Andreev bound states in the quantum spin Hall insulator HgTe},
  author={Bocquillon, Erwann and Deacon, Russell S and Wiedenmann, Jonas and Leubner, Philipp and Klapwijk, Teunis M and Br{\"u}ne, Christoph and Ishibashi, Koji and Buhmann, Hartmut and Molenkamp, Laurens W},
  journal={Nature Nanotechnology},
  volume={12},
  number={2},
  pages={137--143},
  year={2017},
  publisher={Nature Publishing Group UK London}
}

@article{Deacon2017a,
  title = {Josephson Radiation from Gapless Andreev Bound States in HgTe-Based Topological Junctions},
  author = {Deacon, R. S. and Wiedenmann, J. and Bocquillon, E. and Dom\'{\i}nguez, F. and Klapwijk, T. M. and Leubner, P. and Br\"une, C. and Hankiewicz, E. M. and Tarucha, S. and Ishibashi, K. and Buhmann, H. and Molenkamp, L. W.},
  journal = {Phys. Rev. X},
  volume = {7},
  issue = {2},
  pages = {021011},
  numpages = {7},
  year = {2017},
  month = {Apr},
  publisher = {American Physical Society},
  doi = {10.1103/PhysRevX.7.021011},
  url = {https://link.aps.org/doi/10.1103/PhysRevX.7.021011}
}

@article{Nichele2017a,
  title = {Scaling of Majorana Zero-Bias Conductance Peaks},
  author = {Nichele, Fabrizio and Drachmann, Asbj\o{}rn C. C. and Whiticar, Alexander M. and O'Farrell, Eoin C. T. and Suominen, Henri J. and Fornieri, Antonio and Wang, Tian and Gardner, Geoffrey C. and Thomas, Candice and Hatke, Anthony T. and Krogstrup, Peter and Manfra, Michael J. and Flensberg, Karsten and Marcus, Charles M.},
  journal = {Phys. Rev. Lett.},
  volume = {119},
  issue = {13},
  pages = {136803},
  numpages = {5},
  year = {2017},
  month = {Sep},
  publisher = {American Physical Society},
  doi = {10.1103/PhysRevLett.119.136803},
  url = {https://link.aps.org/doi/10.1103/PhysRevLett.119.136803}
}

@article{Fu2008a,
  title = {Superconducting Proximity Effect and Majorana Fermions at the Surface of a Topological Insulator},
  author = {Fu, Liang and Kane, C. L.},
  journal = {Phys. Rev. Lett.},
  volume = {100},
  issue = {9},
  pages = {096407},
  numpages = {4},
  year = {2008},
  month = {Mar},
  publisher = {American Physical Society},
  doi = {10.1103/PhysRevLett.100.096407},
  url = {https://link.aps.org/doi/10.1103/PhysRevLett.100.096407}
}

@article{Fu2009a,
   author    = {Liang Fu and C. L. Kane},
   title     = {Josephson current and noise at a superconductor quantum spin Hall insulator Fsuperconductor junction},
   journal   = {Phys. Rev. B},
   volume    = {79},
   pages     = {161408(R)},
   year      = {2009}
}

@article{Marra2019a,
  title = {Topologically nontrivial Andreev bound states},
  author = {Marra, Pasquale and Nitta, Muneto},
  journal = {Phys. Rev. B},
  volume = {100},
  issue = {22},
  pages = {220502},
  numpages = {7},
  year = {2019},
  month = {Dec},
  publisher = {American Physical Society},
  doi = {10.1103/PhysRevB.100.220502},
  url = {https://link.aps.org/doi/10.1103/PhysRevB.100.220502}
}

@article{Moore2018a,
  title = {Quantized zero-bias conductance plateau in semiconductor-superconductor heterostructures without topological Majorana zero modes},
  author = {Moore, Christopher and Zeng, Chuanchang and Stanescu, Tudor D. and Tewari, Sumanta},
  journal = {Phys. Rev. B},
  volume = {98},
  issue = {15},
  pages = {155314},
  numpages = {6},
  year = {2018},
  month = {Oct},
  publisher = {American Physical Society},
  doi = {10.1103/PhysRevB.98.155314},
  url = {https://link.aps.org/doi/10.1103/PhysRevB.98.155314}
}

@article{Pakizer2021a,
  title = {Signatures of topological transitions in the spin susceptibility of Josephson junctions},
  author = {Pakizer, Joseph D. and Matos-Abiague, Alex},
  journal = {Phys. Rev. B},
  volume = {104},
  issue = {10},
  pages = {L100506},
  numpages = {6},
  year = {2021},
  month = {Sep},
  publisher = {American Physical Society},
  doi = {10.1103/PhysRevB.104.L100506},
  url = {https://link.aps.org/doi/10.1103/PhysRevB.104.L100506}
}

@article{Oladunjoye2019a,
  title = {Supercurrent Detection of Topologically Trivial Zero-Energy States in Nanowire Junctions},
  author = {Awoga, Oladunjoye A. and Cayao, Jorge and Black-Schaffer, Annica M.},
  journal = {Phys. Rev. Lett.},
  volume = {123},
  issue = {11},
  pages = {117001},
  numpages = {7},
  year = {2019},
  month = {Sep},
  publisher = {American Physical Society},
  doi = {10.1103/PhysRevLett.123.117001},
  url = {https://link.aps.org/doi/10.1103/PhysRevLett.123.117001}
}

@article{Ivanov2001a,
  title = {Non-Abelian Statistics of Half-Quantum Vortices in p-Wave Superconductors},
  author = {Ivanov, D. A.},
  journal = {Phys. Rev. Lett.},
  volume = {86},
  issue = {2},
  pages = {268--271},
  numpages = {0},
  year = {2001},
  month = {Jan},
  publisher = {American Physical Society},
  doi = {10.1103/PhysRevLett.86.268},
  url = {https://link.aps.org/doi/10.1103/PhysRevLett.86.268}
}

@article{Kitaev2001a,
  author={A Yu Kitaev},
  title={Unpaired Majorana fermions in quantum wires},
  journal={Physics-Uspekhi},
  volume={44},
  number={10S},
  pages={131},
  url={http://stacks.iop.org/1063-7869/44/i=10S/a=S29},
  year={2001}
}

@article{Liu2017a,
  title = {Andreev bound states versus Majorana bound states in quantum dot-nanowire-superconductor hybrid structures: Trivial versus topological zero-bias conductance peaks},
  author = {Liu, Chun-Xiao and Sau, Jay D. and Stanescu, Tudor D. and Das Sarma, S.},
  journal = {Phys. Rev. B},
  volume = {96},
  issue = {7},
  pages = {075161},
  numpages = {29},
  year = {2017},
  month = {Aug},
  publisher = {American Physical Society},
  doi = {10.1103/PhysRevB.96.075161},
  url = {https://link.aps.org/doi/10.1103/PhysRevB.96.075161}
}

@article{Dmytruk2020a,
  title = {Pinning of Andreev bound states to zero energy in two-dimensional superconductor- semiconductor Rashba heterostructures},
  author = {Dmytruk, Olesia and Loss, Daniel and Klinovaja, Jelena},
  journal = {Phys. Rev. B},
  volume = {102},
  issue = {24},
  pages = {245431},
  numpages = {7},
  year = {2020},
  month = {Dec},
  publisher = {American Physical Society},
  doi = {10.1103/PhysRevB.102.245431},
  url = {https://link.aps.org/doi/10.1103/PhysRevB.102.245431}
}

@article{Majorana1937a,
   author    = {Ettore Majorana},
   title     = {Teoria simmetrica dell'elettrone e del positrone},
   journal   = {Nuovo Cimento},
   year      = {1937},
   volume    = {14},
   doi       = {10.1007/BF02961314},
   pages     = {171}
}

@article{Mourik2012a,
   author    = {V. Mourik and K. Zuo and S. M. Frolov and S. R. Plissard and E. P. A. M. Bakkers and L. P. Kouwenhoven},
   title     = {Signatures of Majorana Fermions in Hybrid Superconductor-Semiconductor Nanowire Devices},
   journal   = {Science},
   year      = {2012},
   volume    = {336},
   doi       = {10.1126/science.1222360},
   pages     = {1003}
}

@article{Li2018a,
  title = {$4\pi$-periodic Andreev bound states in a Dirac semimetal},
  author = {Li, Chuan and de Boer, Jorrit C. and de Ronde, Bob and Ramankutty, Shyama V. and van Heumen, Erik and Huang, Yingkai and de Visser, Anne and Golubov, Alexander A. and Golden, Mark S. and Brinkman, Alexander},
  journal = {Nat. Materials},
  volume = {17},
  issue = {10},
  pages = {880},
  year = {2018},
  publisher = {Nature},
  doi = {10.1038/s41563-018-0158-6},
  url = {https://doi.org/10.1038/s41563-018-0158-6}
}

@article{Read2000a,
  title = {Paired states of fermions in two dimensions with breaking of parity and time-reversal symmetries and the fractional quantum Hall effect},
  author = {Read, N. and Green, Dmitry},
  journal = {Phys. Rev. B},
  volume = {61},
  issue = {15},
  pages = {10267--10297},
  numpages = {0},
  year = {2000},
  month = {Apr},
  publisher = {American Physical Society},
  doi = {10.1103/PhysRevB.61.10267},
  url = {https://link.aps.org/doi/10.1103/PhysRevB.61.10267}
}

@article{Rokhinson2012a,
   author    = {Leonid P. Rokhinson and Xinyu Liu and Jacek K. Furdyna},
   title     = {The fractional a.c. Josephson effect in a semiconductor–superconductor nanowire as a signature of Majorana particles},
   journal   = {Nat. Phys.},
   year      = {2012},
   volume    = {1038},
   doi = {10.1038/nphys2429},
   pages     = {2429}
}

@article{San-Jose2015a,
  title = {Majorana Zero Modes in Graphene},
  author = {San-Jose, P. and Lado, J. L. and Aguado, R. and Guinea, F. and Fern\'andez-Rossier, J.},
  journal = {Phys. Rev. X},
  volume = {5},
  issue = {4},
  pages = {041042},
  numpages = {15},
  year = {2015},
  month = {Dec},
  publisher = {American Physical Society},
  doi = {10.1103/PhysRevX.5.041042}
}

@article{Volovik1999a,
   author    = {G E Volovik},
   title     = {Fermion zero modes on vortices in chiral superconductors},
   journal   = {JETP Lett.},
   year      = {1999},
   volume    = {70},
   doi       = {10.1134/1.568223},
   pages     = {609}
}

@article{Wiedenmann2016a,
	Author = {Wiedenmann, J. and Bocquillon, E. and Deacon, R. S. and Hartinger, S. and Herrmann, O. and Klapwijk, T. M. and Maier, L. and Ames, C. and Bruene, C. and Gould, C. and Oiwa, A. and Ishibashi, K. and Tarucha, S. and Buhmann, H. and Molenkamp, L. W.},
	Journal = {Nature Communications},
	Month = {01},
	Pages = {10303},
	Title = {$4\pi$-periodic Josephson supercurrent in HgTe-based topological Josephson junctions},
	Volume = {7},
	doi = {10.1038/ncomms10303},
	Year = {2016}}

@article{Deng2012a,
author = {Deng, M. T. and Yu, C. L. and Huang, G. Y. and Larsson, M. and Caroff, P. and Xu, H. Q.},
title = {Anomalous Zero-Bias Conductance Peak in a Nb–InSb Nanowire–Nb Hybrid Device},
journal = {Nano Letters},
volume = {12},
number = {12},
doi = {10.1021/nl303758w},
pages = {6414-6419},
year = {2012}
}

@article{Law2009a,
  title = {Majorana Fermion Induced Resonant Andreev Reflection},
  author = {Law, K. T. and Lee, Patrick A. and Ng, T. K.},
  journal = {Phys. Rev. Lett.},
  volume = {103},
  issue = {23},
  pages = {237001},
  numpages = {4},
  year = {2009},
  month = {Dec},
  publisher = {American Physical Society},
  doi = {10.1103/PhysRevLett.103.237001},
  url = {https://link.aps.org/doi/10.1103/PhysRevLett.103.237001}
}

@article{Bauer2021a,
  title = {Quantized phase-coherent heat transport of counterpropagating Majorana modes},
  author = {Bauer, Alexander G. and Scharf, Benedikt and Molenkamp, Laurens W. and Hankiewicz, Ewelina M. and Sothmann, Bj\"orn},
  journal = {Phys. Rev. B},
  volume = {104},
  issue = {20},
  pages = {L201410},
  numpages = {6},
  year = {2021},
  month = {Nov},
  publisher = {American Physical Society},
  doi = {10.1103/PhysRevB.104.L201410},
  url = {https://link.aps.org/doi/10.1103/PhysRevB.104.L201410}
}

@ARTICLE{Laroche2019a,
       author = {Laroche, Dominique and Bouman, Dani\"el and van Woerkom, David J. and Proutski, Alex and Murthy, Chaitanya and Pikulin, Dmitry I. and Nayak, Chetan and van Gulik, Ruben J. J. and Nygard, Jesper and Krogstrup, Peter and Kouwenhoven, Leo P. and Geresdi, Attila},
        title = {Observation of the $4\pi$-periodic Josephson effect in indium arsenide nanowires},
      journal = {Nat. Comm.},
        volume = {10},
         year = {2019},
        pages = {245},
   doi = {10.1038/s41467-018-08161-2},
  url = {https://doi.org/10.1038/s41467-018-08161-2}
}

@article{Beenakker2006a,
  title = {Specular Andreev Reflection in Graphene},
  author = {Beenakker, C. W. J.},
  journal = {Phys. Rev. Lett.},
  volume = {97},
  issue = {6},
  pages = {067007},
  numpages = {4},
  year = {2006},
  month = {Aug},
  publisher = {American Physical Society},
  doi = {10.1103/PhysRevLett.97.067007},
  url = {https://link.aps.org/doi/10.1103/PhysRevLett.97.067007}
}

@article{Lopez2014a,
  title = {Thermoelectrical detection of Majorana states},
  author = {L\'opez, Rosa and Lee, Minchul and Serra, Lloren\ifmmode \mbox{\c{c}}\else \c{c}\fi{} and Lim, Jong Soo},
  journal = {Phys. Rev. B},
  volume = {89},
  issue = {20},
  pages = {205418},
  numpages = {7},
  year = {2014},
  month = {May},
  publisher = {American Physical Society},
  doi = {10.1103/PhysRevB.89.205418},
  url = {https://link.aps.org/doi/10.1103/PhysRevB.89.205418}
}

@article{Sturm2025a,
  title = {Transport signatures of inverted Andreev bands in topological Josephson junctions},
  author = {Sturm, Jonathan and Klees, Raffael L. and Hankiewicz, Ewelina M. and Gresta, Daniel},
  journal = {Phys. Rev. B},
  volume = {111},
  issue = {18},
  pages = {184518},
  numpages = {12},
  year = {2025},
  month = {May},
  publisher = {American Physical Society},
  doi = {10.1103/PhysRevB.111.184518},
  url = {https://link.aps.org/doi/10.1103/PhysRevB.111.184518}
}

@article{Wang2011a,
  title = {Topological field theory and thermal responses of interacting topological superconductors},
  author = {Wang, Zhong and Qi, Xiao-Liang and Zhang, Shou-Cheng},
  journal = {Phys. Rev. B},
  volume = {84},
  issue = {1},
  pages = {014527},
  numpages = {5},
  year = {2011},
  month = {Jul},
  publisher = {American Physical Society},
  doi = {10.1103/PhysRevB.84.014527},
  url = {https://link.aps.org/doi/10.1103/PhysRevB.84.014527}
}

@article{Akhmerov2011a,
  title = {Quantized Conductance at the Majorana Phase Transition in a Disordered Superconducting Wire},
  author = {Akhmerov, A. R. and Dahlhaus, J. P. and Hassler, F. and Wimmer, M. and Beenakker, C. W. J.},
  journal = {Phys. Rev. Lett.},
  volume = {106},
  issue = {5},
  pages = {057001},
  numpages = {4},
  year = {2011},
  month = {Jan},
  publisher = {American Physical Society},
  doi = {10.1103/PhysRevLett.106.057001},
  url = {https://link.aps.org/doi/10.1103/PhysRevLett.106.057001}
}

@article{Ribeiro_2022,
  title = {Spin-polarized Majorana zero modes in double zigzag honeycomb nanoribbons},
  author = {Ribeiro, R. C. Bento and Correa, J. H. and Ricco, L. S. and Seridonio, A. C. and Figueira, M. S.},
  journal = {Phys. Rev. B},
  volume = {105},
  issue = {20},
  pages = {205115},
  numpages = {11},
  year = {2022},
  month = {May},
  publisher = {American Physical Society},
  doi = {10.1103/PhysRevB.105.205115}
}

@Article{Ribeiro_2023,
    author={Bento Ribeiro, R. C.
    and Correa, J. H.
    and Ricco, L. S.
    and Shelykh, I. A.
    and Continentino, Mucio A.
    and Seridonio, A. C.
    and Minissale, M.
    and Le Lay, G.
    and Figueira, M. S.},
    title={Spin-polarized Majorana zero modes in proximitized superconducting penta-silicene nanoribbons},
    journal={Scientific Reports},
    year={2023},
    month={Oct},
    day={20},
    volume={13},
    number={1},
    pages={17965},
    issn={2045-2322},
    doi={10.1038/s41598-023-44739-7}
}

@article{Perge2014a,
    author = {Stevan Nadj-Perge  and Ilya K. Drozdov  and Jian Li  and Hua Chen  and Sangjun Jeon  and Jungpil Seo  and Allan H. MacDonald  and B. Andrei Bernevig  and Ali Yazdani },
    title = {Observation of Majorana fermions in ferromagnetic atomic chains on a superconductor},
    journal = {Science},
    volume = {346},
    number = {6209},
    pages = {602-607},
    year = {2014},
    doi = {10.1126/science.1259327},
    eprint = {https://www.science.org/doi/pdf/10.1126/science.1259327}
}

@article{Jeon2017a,
    author = {Sangjun Jeon  and Yonglong Xie  and Jian Li  and Zhijun Wang  and B. Andrei Bernevig  and Ali Yazdani },
    title = {Distinguishing a Majorana zero mode using spin-resolved measurements},
    journal = {Science},
    volume = {358},
    number = {6364},
    pages = {772-776},
    year = {2017},
    doi = {10.1126/science.aan3670},
    eprint = {https://www.science.org/doi/pdf/10.1126/science.aan3670}
}

@Article{Penaranda2023a,
	title={{Majorana bound states in encapsulated bilayer graphene}},
	author={Fernando Peñaranda and Ramón Aguado and Elsa Prada and Pablo San-Jose},
	journal={SciPost Phys.},
	volume={14},
	pages={075},
	year={2023},
	publisher={SciPost},
	doi={10.21468/SciPostPhys.14.4.075}
}

@article{Suominen2017a,
  title = {Zero-Energy Modes from Coalescing Andreev States in a Two-Dimensional Semiconductor-Superconductor Hybrid Platform},
  author = {Suominen, H. J. and Kjaergaard, M. and Hamilton, A. R. and Shabani, J. and Palmstr\o{}m, C. J. and Marcus, C. M. and Nichele, F.},
  journal = {Phys. Rev. Lett.},
  volume = {119},
  issue = {17},
  pages = {176805},
  numpages = {5},
  year = {2017},
  month = {Oct},
  publisher = {American Physical Society},
  doi = {10.1103/PhysRevLett.119.176805}
}

@article{Ning-Xuan2022a,
  title = {Half-integer quantized thermal conductance plateau in chiral topological superconductor systems},
  author = {Yang, Ning-Xuan and Yan, Qing and Sun, Qing-Feng},
  journal = {Phys. Rev. B},
  volume = {105},
  issue = {12},
  pages = {125414},
  numpages = {8},
  year = {2022},
  month = {Mar},
  publisher = {American Physical Society},
  doi = {10.1103/PhysRevB.105.125414},
  url = {https://link.aps.org/doi/10.1103/PhysRevB.105.125414}
}

@article{Liu2025,
  title = {Period-doubling in the phase dynamics of a shunted HgTe quantum well Josephson junction},
  author = {Liu, Wei and Piatrusha, Stanislau U. and Liang, Xianhu
 and Upadhyay, Sandeep and Fürst, Lena and Gould, Charles
and Kleinlein, Johannes and Buhmann, Hartmut and Stehno, Martin P. and Molenkamp, Laurens W.},
  journal = {Nat. Comm.},
  volume = {16},
  issue = {1},
  pages = {3068},
  year = {2025},
  doi = {10.1038/s41467-025-58299-z},
  url = {https://doi.org/10.1038/s41467-025-58299-z}
}

@article{Sticlet2018a,
  title = {Dissipation-enabled fractional Josephson effect},
  author = {Sticlet, Doru and Sau, Jay D. and Akhmerov, Anton},
  journal = {Phys. Rev. B},
  volume = {98},
  issue = {12},
  pages = {125124},
  numpages = {16},
  year = {2018},
  month = {Sep},
  publisher = {American Physical Society},
  doi = {10.1103/PhysRevB.98.125124},
  url = {https://link.aps.org/doi/10.1103/PhysRevB.98.125124}
}

\end{document}